\documentclass[a4paper,12pt]{article}

\usepackage{amssymb,amsmath,mathbbol,mathrsfs}
\usepackage[usenames,dvipsnames]{color}
\usepackage{hyperref}
\usepackage{xcolor}
\usepackage{stmaryrd}
\usepackage{authblk}
\usepackage{framed}
\usepackage{empheq}
\usepackage{slashed}
\usepackage{hyphenat}
\usepackage{cite}
\usepackage{apacite}
\usepackage{natbib}
\usepackage[normalem]{ulem} 
\usepackage{soul}

\usepackage{marginnote}    

\usepackage[left=.8in,right=.8in,top=.8in,bottom=.8in]{geometry}                  

\usepackage{sectsty}
\allsectionsfont{\sffamily\mdseries\upshape} 
\usepackage{tocloft}

\makeatletter
\renewenvironment{abstract}{%
    \if@twocolumn
      \section*{\abstractname}%
    \else 
      \begin{center}%
        {\bfseries\sffamily\abstractname\vspace{\z@}}
      \end{center}%
      \quotation
    \fi}
    {\if@twocolumn\else\endquotation\fi}
\makeatother

\numberwithin{equation}{section}
5

\hypersetup{
	colorlinks=true,         
	linkcolor=MidnightBlue,          
	citecolor=BrickRed,        
	urlcolor=MidnightBlue            
}

\newcommand\mathC{\mkern1mu\raise2.2pt\hbox{$\scriptscriptstyle|$}
        {\mkern-7mu\rm C}}              

\newcommand{\R}{{\mathbb{R}}}

\newcommand{\be}{\begin{equation}}
\newcommand{\ee}{\end{equation}}

\renewcommand{\d}{{\mathrm{d}}}
\newcommand{\D}{{\mathrm{D}}}

\newcommand{\pp}{{\partial}}

\renewcommand{\bar}{\overline}
\newcommand{\dd}{{\mathbb{d}}}

\newcommand{\La}{\mathcal{L}}

\newcommand{\cint}{{\int\kern-.87em{<}}}
\newcommand{\sint}{{\int\kern-.75em{\sim}}}
\newcommand{\fint}{{\int\kern-1.00em{\int}}}

\newcommand{\bb}{\mathbb}

\let\oldmarginpar\marginpar
\renewcommand\marginpar[1]{\oldmarginpar{\color{red}\raggedright\footnotesize #1}}

\newcommand{\jb}{\color{ForestGreen}}

\newcommand{\old}{\color{red}}

\begin{document}

\title{How to Choose a Gauge? The case of Hamiltonian Electromagnetism}
\author{Henrique Gomes and Jeremy Butterfield\footnote{\href{mailto:gomes.ha@gmail.com}{gomes.ha@gmail.com, jb56@cam.ac.uk}} \\\it Trinity College, University of Cambridge }

\maketitle
\vspace{-1cm}
\begin{abstract}
We develop some ideas about gauge symmetry in the context of Maxwell's theory of electromagnetism in the Hamiltonian formalism. One great benefit of this  formalism is that it pairs momentum and   configurational 
 degrees of freedom, so that a decomposition of one side into subsets can be translated into a decomposition of the other.  In the case of electromagnetism, this enables us to pair degrees of freedom of the electric field with degrees of freedom of the vector potential. Another benefit is that the formalism algorithmically identifies  subsets of the equations of motion that represent   time-dependent symmetries. For electromagnetism, these two benefits allow us to define gauge-fixing  in parallel  to special decompositions of the electric field.  More specifically, we apply  the Helmholtz decomposition theorem to split the electric field into its Coulombic and radiative parts, and show how this gives a special role to the Coulomb gauge (i.e. div$({\bf A}) = 0$).  We relate this argument to Maudlin's (2018) discussion, which advocated the Coulomb gauge. 
 \end{abstract}

\tableofcontents

 \section{Introduction}\label{intro}

Recently,   \cite{Maudlin_ontology} has exhorted not only philosophers, but also modern theoretical physicists, to be clearer about their theories' ontology.  
He writes: `both the glory and the bane of modern physics is its highly mathematical character. This has provided both for the calculation of stunningly precise
predictions and for the endemic unclarity about the physical ontology being postulated'  \citet[p.5]{Maudlin_ontology}. We agree. Today, more than in yesteryear, a theory's mathematical formalism is often interpreted \emph{en bloc}. No special care is taken to specify: which parts represent ontology, `what there is' (and within that: what is basic or fundamental, and what derived or composite); and which parts represent `how it behaves' (which Maudlin (p. 4) calls `nomology': in particular, dynamics); and which parts represent nothing physical, but instead mathematics (which, though unphysical, can of course be invaluable for calculation). 

We also endorse Maudlin's programme to develop presentations of our physical theories that are clear about these distinctions; (though one should of course accept  that what we usually consider to be one theory may admit two such meritorious presentations---no uniqueness claim is required.)\footnote{So this programme is consistent, in particular, with rejecting the logical positivists’ aim of presenting physical theories with a once-for-all division of fact and convention: a rejection we share with e.g. \cite{Putnam_analytic}.}
  

As a case study of his programme, Maudlin considers a theory that, in the hierarchy of mathematical sophistication of modern physics, sits relatively low, viz. classical electromagnetism. He then applies the results to assess some proposed interpretations of the Aharonov-Bohm effect. 

In this paper, we also will consider classical electromagnetism,  and with an overall aim similar to Maudlin's---to clarify interpretative issues. But we focus on the Hamiltonian formulation of the theory  (Section  \ref{sec:HamEM}): which we lead up to, by first expounding the Hamiltonian framework  for finite-dimensional systems, rather than for field theories (Section  \ref{sec:Ham}). This exposition will include comparison with the Lagrangian framework, and the treatment of constraints. Despite our adopting the Hamiltonian framework, some of our conclusions, technical as well as interpretative, will be concordant with Maudlin’s: in particular, about the Coulomb gauge having a special role.

It is often remarked that the Hamiltonian framework’s use of a time parameter, and-or of a `3+1’ split of spacetime, carries the price, for special relativistic theories like electromagnetism, that one loses manifest Lorentz invariance . We of course accept that this is a limitation; though we note that often (including in our discussion below) the Lagrangian framework also uses a time parameter. 

But  for interpreting the gauge aspects of electromagnetism, the Hamiltonian framework has two 
  countervailing benefits. 
 First: its pairing of momentum and configurational  degrees of freedom means that we can  pair degrees of freedom of the electric field with degrees of freedom of the vector potential. Second: the Hamiltonian framework illuminates symmetries that are time-dependent of the sort associated (in both it and the Lagrangian framework), with constraints and the failure of determinism.  This illumination comes from the way the Hamiltonian framework algorithmically identifies  subsets of the equations of motion that represent such symmetries   (viz. `the Dirac algorithm'). Combining these benefits gives an illuminating, and interpretatively clear,  splitting of the  electric field into its Coulombic and radiative parts. Besides, this splitting gives a special role to the Coulomb gauge,  i.e. div$({\bf A}) = 0$.\footnote{ Maudlin also advocates this gauge, but for very different reasons than us. We discuss the differences in Sections \ref{sec:Gauss} and \ref{concl} (cf. also footnote \ref{analo}), but in short: he makes a controversial ontological proposal, while we, less contentiously, see the gauge as ``merely’’ natural, because of its reflecting the splitting of the electric field.}


 So our plan is as follows. In Section  \ref{sec:Ham}, we will present the 
 features of the Hamiltonian  framework that will be useful for us. We will first describe how constraints emerge from the Legendre transform of the Lagrangian, and the implications for (in)determinism. Since constraints will play an important role in the paper, we will spend much of the Section justifying their special status amongst the equations that specify a theory.  In particular, we will see that---although constraints imply that, roughly speaking, many velocities $\dot q$ in the Lagrangian framework are mapped by the Legendre transformation to one momentum $p$ in the Hamiltonian framework---nevertheless, under reasonable conditions (viz. the constraints being first-class and so generating symmetries), the gauge orbits in the constraint surface within the Hamiltonian framework correspond one-to-one  to sub-manifolds of the Lagrangian state space: (cf. the two paragraphs after equation \ref{eq:Ham_Gamma}).


  In Section \ref{sec:HamEM} we will apply these ideas to electromagnetism: the Gauss constraint emerges from the Dirac algorithm, and it is the generator of gauge symmetries. Finally, we will split the electric field into the part that is uniquely fixed by the Gauss constraint---and so by the instantaneous distribution of charges---and the remainder. We thereby find a symplectically corresponding decomposition of the gauge potential into a part that is pure gauge and a remainder: call them $X$ and $Y$, respectively. If we single out the  scalar degree of freedom of the electric field that  is fixed by the Gauss constraint, by representing it as a gradient, it turns out that, on the other side of the symplectic correspondence,  Coulomb gauge is singled out. More precisely: if we demand that in our split of the gauge potential, the part  $Y$ of the gauge potential that remains after we extract the pure gauge part    $X$, 
  is dynamically independent from the part of the electric field that is fixed by the Gauss constraint,  then this non-gauge part  $Y$  will obey the Coulomb gauge equation.  
  
 In other, somewhat less technical, words: our main idea is as follows. The Coulombic part of the electric field is its electrostatic-like component, which is determined by the instantaneous distribution of charges. The electric field has the gauge potential as its conjugate. And as to the Coulombic part of the electric field, there is a part of the gauge potential that is not at all conjugate to it (i.e. is symplectically orthogonal to it). That part of the potential satisfies the Coulomb gauge condition.\footnote{So the nomenclature is confusing, since the Coulombic part of the electric field corresponds to, i.e. is conjugate to, the {\em other} part of the gauge potential than that which satisfies the Coulomb gauge condition. Thus one might facetiously propose that for clarity, we should: (i) associate the adjective `Coulombic’ with Coulomb’s historical work on electrostatics, and so apply it only to the electrostatic-like component of the electric field; and (ii) as regards the gauge potential,  re-name the gauge condition as the {\em `ortho-Coulomb condition’}.} Besides, this result is worth expounding, since one usually thinks of a choice of gauge as motivated by calculational convenience, often for a specific problem---and so from a general theoretical perspective, as completely arbitrary: whereas this result shows that the choice is related  to a physically natural, and general, splitting of the electric field.

In the short final Section \ref{concl}, we conclude; and sketch how this work can be set against the background, not just of Maudlin's discussion, but of other similar enterprises,   viz.  those pursued in \cite{Samediff_1, Samediff_2, Gomes_new, GomesNoether, GomesButterfieldRoberts, DES_gf}. 

We now end this introduction by presenting three overall morals, especially in relation to the Aharonov-Bohm effect. 

\subsection{Three morals}\label{decide!}
 Elsewhere one of us  \cite[Sec. 3]{Samediff_1},  has urged an interpretation of gauge theories---more specifically: the family of Abelian and non-Abelian classical Yang-Mills theories, which includes electromagnetism---that takes a single physical possibility to correspond to an entire gauge-equivalence class.\footnote{\label{analo}{So this is analogous to what is often called `sophisticated substantivalism' in the philosophy of general relativity; (cf. footnote \ref{gr}). It  is also disanalogous to Maudlin, since in the enterprise of interpreting gauge, he investigates ontology associated with different representatives of each equivalence class. So his reasons for advocating the Coulomb gauge are very different from our reasons for 
 seeing it as special.}} In Maudlin's jargon of ontology (`what there is') and nomology (`how it behaves'), this interpretation, in brief,  is that:\\
 \indent  \indent (i) \emph{the ontology} consists of a field over spacetime, that encodes the relations  between charges that interact via a particular type of force (e.g. electromagnetism), i.e. the field encodes  a relation of sameness of charges across  spacetime called `parallel transport';\footnote{This is argued by  \citet[Sec. 3]{Samediff_1} to bear a straightforward analogy to the more familiar parallel transport of vectors on spacetime, which is fixed by the metric.} \\
 \indent  \indent (ii) \emph{ the nomology} describes how this sameness relation is constrained by the distributions of charges in spacetime. \\
 Besides: although this interpretation was articulated in the Lagrangian (covariant) framework, we see no significant obstacle to an appropriate translation to the Hamiltonian framework (though  there are some subtleties  about how  to do this: cf.  the end of Section \ref{sec:Ham_symp} and \citep[Sec. 11]{Belot2003}).
 Thus   one main  aim of this paper is to  give such a translation.     

 In either  framework,  this summary, (i) and (ii), is of course a ``high altitude'' view of the theory. Nonetheless,  we  can already see how it accommodates facts that are characterised only non-locally  in terms of their spacetime properties. Thus, for example, one can parallel transport an internal quantity around a loop in spacetime, and whether that quantity comes back to its original value or not can depend on facts {\em outside} the vicinity of the loop \cite[Sec. 2.3]{Samediff_2}.\footnote{\label{vicin}{In more detail: Suppose we take all the \emph{local} gauge-invariant facts to be given/determined. This includes a field called \emph{curvature} in the Abelian theory (and similarly, traces of products of curvature in the non-Abelian theory). Then suppose we have two possible worlds, $w$ and $w'$, and we can identify a loop in spacetime in $w$ with one in $w'$. Even if $w$ and $w'$ agree about the distribution of the local gauge-invariant quantities in the vicinity of the loop, they may disagree about  whether the  original value of the charges will coincide with their values after parallel transport around the loop. This fact thus depends on some local gauge-invariant quantities \emph{outside} the vicinity of the loop.  Compare the discussion of holonomies, at the end of this Section and in the Appendix.}}
  
  In the standard jargon of electromagnetism, this type of non-locality is associated with  the Aharonov-Bohm effect and the introduction of \emph{gauge} and {\em vector potentials}.  The most important lesson of the effect is   of course that there is physical significance in gauge. This is highlighted already in the opening of the original paper   \citep{aharonovbohm1959}:
\begin{quote}
In classical electrodynamics, the vector and scalar potentials were first introduced as a
convenient mathematical aid for calculating the fields. It is true that in order to obtain a
classical canonical formalism, the potentials are needed. Nevertheless, the fundamental
equations of motion can always be expressed directly in terms of the fields alone.
In the quantum mechanics, however, the canonical formalism is necessary, and as a result,
the potentials cannot be eliminated from the basic equations. Nevertheless, these equations,
as well as the physical quantities, are all gauge invariant; so that it may seem that even in
quantum mechanics, the potentials themselves have no independent significance.
In this paper we shall show that the above conclusions are not correct and that a further
interpretation of the potentials is needed in quantum mechanics. 
\end{quote}

But we would like to stress  two points of clarification about  this quotation.  They will lead to three ``morals’’   which we will announce here, but present in detail in Sections \ref{sec:Ham} and  \ref{sec:HamEM}. \\

(1): First: although the potentials appear prominently in the theory, we need not interpret as physically different, two potentials that are related by a certain type of transformation, labeled \emph{gauge transformation}.  Thus the  `independent significance' of the gauge potentials needs to be taken {\em modulo} such transformations. 
 
 The lack of physical difference between potentials related by gauge transformations  is incorporated in the  interpretation that we endorse, glimpsed above from a high altitude---and usually articulated mathematically  in terms of fiber bundles. This yields a precise sense in which it is only the \emph{equivalence class} of potentials under gauge transformations that encodes the  parallel transport of internal quantities. Thus in Yang-Mills theories, gauge-related models correspond to the same  physical situation.\footnote{\label{gr}{So this is analogous to saying, about general relativity, that diffeomorphism-related models of the metric correspond to the same  physical possibility: an interpretation that we argue (\cite{GomesButterfield_hole}) is recommended, though not mandated, by the mathematics. As argued at length in \citep{Samediff_1, Samediff_2}, there is thus no major interpretative difference between the gauge transformations of Yang-Mills theories and the diffeomorphisms of general relativity.}} 
 
  If one rejects this first point, then instead of articulating an ontology for the equivalence class of gauge potentials, one would try to articulate ontologies for different choices of an element of the class: i.e. for different choices of `gauge-fixing'. As we mentioned in footnote \ref{analo}, this is \cite{Maudlin_ontology}'s approach; (cf. also \cite[Sections 8-9]{Mulder_AB}).  In the end, Maudlin settles on Coulomb gauge as 
   combining a relativistic nomology with an ontology that requires a preferred foliation of spacetime: and is thus friendly to the pilot-wave/Bohmian   interpretations of quantum mechanics that he favours.   \\
  
(2): Our second  point of clarification is that, although the  experimental set-ups implemented hitherto for observing the Aharonov-Bohm effect have a quantum component---viz. an effect on the phase of a quantum wave-function---the {\em classical} theory already interprets the situations that give rise to the effect as physically distinct. 
The reason---in short, and in an elementary 3-vector formalism---lies in the facts that: \\
\indent \indent (i) the gauge transformations for the vector potential are ${\bf A} \mapsto {\bf A} + \nabla \lambda$, for $\lambda$ any scalar; \\
\indent \indent (ii) the magnetic field ${\bf B} \equiv \nabla \wedge {\bf A}$ is unchanged by the addition of any curl-free vector to $\mathbf{A}$, i.e. ${\bf C}$ such that  $\nabla \wedge {\bf C} = 0$; while \\
\indent \indent  (iii) in a non-simply connected space $\nabla \wedge {\bf C} = 0$ does {\em not} imply that ${\bf C} = \nabla \lambda$ for some scalar $\lambda$. (Point (iii)  is often expressed in the language of forms as: a closed form need not be exact.) \\
So the gauge-equivalence classes of  the vector potential $\bf A$ are narrower than the equivalence classes given by addition of a curl-free vector ${\bf C}$. So the classical theory discriminates as physically distinct two configurations (field distributions) of $\bf A$ that (a) do not differ by a gradient $\nabla \lambda$ (are not gauge-equivalent), but that (b) {\em do} differ by a curl-free vector field $\bf C$ (and so determine the same magnetic field  ${\bf B} \equiv \nabla \wedge {\bf A}$). We agree that hitherto no {\em classical} way to experimentally manifest this distinction has been found. But nor is there a no-go theorem vetoing such a classical manifestation: that is, a measurement (ultimately, a pointer-reading) that uses some classical probe to register which of two $\bf A$-configurations, that are physically distinct since not differing merely by a gradient $\nabla \lambda$, but that determine the same magnetic field, is realized.

\medskip

Both these two points of clarification  (especially of course, the first) are  recognised in the literature.\footnote{The second is  well expressed by  \citet[pp. 532-533, pp. 550-554]{Belot1998}, who also addresses judiciously the question of what an interpretation of classical electromagnetism can tell us about a quantum world. A related question concerns what  we can learn about the world from the sort of idealized descriptions usually given of the Aharonov-Bohm effect: including, for example, excluding from space the region where the local gauge-invariant quantities such as curvature (cf. footnote \ref {vicin}) are distinct, i.e. excluding the solenoid. For recent controversy about such idealizations, cf. \cite{ShechAB, Earman2019, DoughertyAB}.




} 
 What matters for us is that the second point is related  to   classical electromagnetism's exhibiting  a  type of \emph{non-locality}. This is usually expressed in terms of holonomies; (though a complete treatment without holonomies can be given---see \cite[Sec. 6]{GomesRiello_new}, \cite{Gomes_new}). Thus \cite{Myrvold2010} notices that, for classical electromagnetism {\em in vacuo}, and for a simply connected region: one can use the composition laws of the holonomies to decompose any gauge-invariant function on  this region into gauge-invariant functions of its component---i.e. mutually exclusive, jointly exhaustive---sub-regions. But, Myrvold continues: for a region that is not simply connected, there are certain ``large'' holonomies that cannot be obtained from composition of holonomies confined to the sub-regions. That is, there are certain global gauge-invariant functions that are not \emph{separable}.  Myrvold concludes that this type of non-locality only arises for non-simply connected regions (and is exhibited by the Aharonov-Bohm effect). 
 
At first sight, this conclusion is in tension with our own discussion, in Section \ref{sec:HamEM} , of the role in the Hamiltonian framework of the {\em Gauss constraint}, i.e. the equation, in elementary terms, that the divergence of $\bf E$ is equal to the charge density $\rho$ (and so vanishes in vacuum, which is the case we shall often focus on): $\nabla \cdot \bf E = \rho$. The starting-point in our discussion of this equation will be the oft-noted fact that it  involves a sort of {\em non-locality}. For it implies that by simultaneously measuring the electric field flux on all of a large  surface surrounding a charge distribution, and integrating, we can ascertain the total amount of charge inside the sphere {\em at the given instant}.  So this non-locality is classical, and regardless of whether the volume enclosing the charge distribution is simply connected  (the surface surrounding the charge distribution could be topologically a doughnut, not a sphere)---which is apparently, at odds with Myrvold's conclusion.      
But here we should recall that Myrvold's analysis is restricted to electromagnetism {\em in vacuo}; and fortunately, when we remove the restriction, the tension disappears. That is: when we allow for charges, the same  line of argument  that Myrvold uses also proves that there are gauge-invariant functions that are not separable---even for simply connected regions.  In short: in the presence of charges, we have  non-locality (and non-separability) even for simply-connected regions.  The Appendix will give more details.

\medskip

We can sum up this discussion of our relation to Maudlin’s enterprise, and of the Aharonov-Bohm effect, in three morals, as follows:\\
\indent \indent (1): Even setting aside Aharonov-Bohm phases, {\em classical} electromagnetism in the Hamiltonian formalism shows a certain kind of (non-signalling) non-locality, namely in the Gauss constraint.\\
\indent \indent (2):   There is no   unique physically preferred split of the field's degrees of freedom into purely local and purely non-local ones. But each such choice of split can be made to correspond to a choice of gauge-fixing. \\
\indent \indent (3): In particular, the Coulomb gauge corresponds in this way to a natural choice of splitting of the electric field; (which, incidentally, buttresses some of \cite{Maudlin_ontology}'s arguments). That is: this gauge follows naturally   from considering the Gauss constraint to single out a `scalar' part  of the electric field that is determined by the instantaneous distribution of charges.


 \section{The Hamiltonian framework}\label{sec:Ham}
 

 One of the most distinctive features of the Hamiltonian framework  is the fact that  the Dirac analysis of constraints provides an algorithm for discovering gauge   symmetries. Once the algorithm succeeds, the equations of motion of a theory are divided into ones that come from constraints and those that we normally think of as generating dynamics.

Since the formalism is slightly unfamiliar to philosophers of physics,  in Section \ref{sec:Lag} we will first review, for  the  case of a non-relativistic mechanical system,   how under-determination of motion i.e. indeterminism can arise, in the {\em Lagrangian} framework. Then we will relate this to the constraints in the Hamiltonian framework that are obtained by a Legendre transformation. Then, in Section \ref{sec:Ham_symp} we will introduce the analysis of symmetry using symplectic geometry.  In these two Subsections, we will, broadly speaking, keep the relation between the Lagrangian and Hamiltonian frameworks simple---though sufficient for  our points about  the case of electromagnetism---by certain restrictions of scope about types of constraint. (These restrictions are summarised in the two paragraphs after equation \ref{eq:Ham_Gamma}.)  Thus we will see that both the Lagrangian and Hamiltonian frameworks show the special status of constraints: they are the generators of symmetries and must be imposed prior to any equation of motion, if there is to be a correspondence between the frameworks.  We end the Section with some philosophical remarks about the status of states that are not in the constraint surface.

\subsection{The Lagrangian framework and indeterminism }\label{sec:Lag}
In the Lagrangian formulation of mechanics, the focus is on curves in the space of all possible instantaneous configurations of the system. For the mechanics of $N$ point-particles parametrized by $\alpha=1, \cdots, N$, let this configuration space be called $\mathcal{Q}$. A \emph{Lagrangian} is a map from the tangent bundle on configuration space to the reals: $\mathcal{L}:T\mathcal{Q}\rightarrow \bb R$. Once integrated along a curve, the Lagrangian yields the \emph{action functional} as a function from curves $\gamma$ in $\mathcal{Q}$ into the reals:
\be S(\gamma):=\int_\gamma \d t\, \mathcal{L}(q^\alpha(t), \dot q^\alpha(t)).\ee 
One then obtains, from the least action
principle $\delta  S=0$, the Euler--Lagrange equations:
\be \label{equ:Euler-Lagrange}\frac{d}{dt}\frac{\pp  \mathcal{L}}{\pp \dot q^\alpha}=\frac{\pp  \mathcal{L}}{\pp  q^\alpha}.
\ee
If we use the chain rule for the $\frac{d}{dt}$ derivative, we get from \eqref{equ:Euler-Lagrange}:
\be\label{equ:accel_lagrange} \ddot q^{\beta}\frac{\pp ^2 \La}{\pp \dot q^{\beta}\pp \dot q^\alpha}+\dot q^{\beta}\frac{\pp ^2 \La}{\pp  q^{\beta}\pp \dot q^\alpha} =\frac{\pp \La}{\pp  q^\alpha}
.\ee
The accelerations are uniquely determined by the positions and velocities if we can isolate $\ddot q^{\beta}$ in this equation. A necessary and sufficient condition for this  is that the matrix $M_{\alpha\beta}:=\frac{\pp ^2 \La}{\pp \dot q^{\beta}\pp \dot q^\alpha} $ be  invertible. If it is not,   the accelerations are undetermined, so that the motion is under-determined by the initial positions and velocities: there is indeterminism at the level of the $2N$ variables, $q^\alpha(t), \dot q^\alpha(t)$.   Assuming we believe that in physical terms, the motion is indeed determined, this indicates a  redundancy in our description of the system. For a philosophical introduction to this indeterminism and redundancy, from a Lagrangian treatment, compare \cite{Wallace_LagSym}.  

Let us see how this redundancy appears in the Hamiltonian formalism. The idea of the {\em Legendre transformation} is that at any point $q \in {\cal Q}$, the Lagrangian $\cal L$ determines a map $\mathrm{Leg}_q$ from the tangent space $T_q{\cal Q}$ at $q$ to its dual space $T^*_q{\cal Q}$. Intuitively speaking, this is the transition ${\dot q} \mapsto p$. To be precise, $\mathrm{Leg}_q$ is defined by
\be
\mathrm{Leg}_q : w = {\dot q}^{\alpha} \frac{\partial }{\partial q^{\alpha}} \in T_q{\cal Q} \; \mapsto \; \frac{\partial {\cal L}}{\partial {\dot q}^{\alpha}} \in T^*_q{\cal Q}  \; .
\ee         
One easily checks that because the canonical momenta $p_\alpha:=\frac{\pp  \La}{\pp  \dot q^\alpha}$ are 1-forms, this definition is coordinate-independent. An equivalent definition, manifestly coordinate-independent and given for all $q \in {\cal Q}$, is the Legendre transformation, $\mathrm{Leg}:T\mathcal{Q}\rightarrow T^*\mathcal{Q}$, defined by
\be\label{eq:Leg}
\forall q \in {\cal Q}, \, \forall v,w \in T_q\mathcal{Q} \; : \; \;  \mathrm{Leg}(w)(v):=\left.\frac{d}{dt}\right|_{t=0}\mathcal{L}(w + t v).
\ee
(Here we take $w,v$ to encode the identity of the base-point $q$, so as to simplify notation, writing $\mathrm{Leg}(w)$ rather than $\mathrm{Leg}(q,w)$ etc.) That is: $\mathrm{Leg}(w)(v)$ is the derivative of $\cal L$ at $w$, along the fibre $T_q{\cal Q}$ of the fibre bundle $T{\cal Q}$, in the direction $v$. Thus $\mathrm{Leg}$ is also called the {\em fibre derivative}. 

Given $\cal L$, we define its energy function $E:T{\cal Q} \rightarrow {\bb R}$ by 
\be\label{Lenerg}
\forall v \in T{\cal Q}, \; E(v) := \mathrm{Leg}(v)(v) \, – \, {\cal L}(v) \; ;
\ee
or in coordinates, 
\be\label{Lenerg2}
E(q^{\alpha}, {\dot q}^{\alpha}) := \frac{\partial {\cal L}}{\partial {\dot q}^{\alpha}}{\dot q}^{\alpha} \, - \, {\cal L}( q^{\alpha}, {\dot q}^{\alpha}).
\ee
Then one shows that 
 $E \circ (\mathrm{Leg})^{-1}$ is, as one would hope, the familiar Hamiltonian function $H: T^*{\cal Q} \rightarrow {\bb R}$ given by $(q,p) \equiv (q^{\alpha}, p_{\alpha}) \mapsto {\dot q}p - {\cal L} \equiv {\dot q}^{\alpha}p_{\alpha} - {\cal L}$, with each ${\dot q}^{\alpha}$ being a function of the $q$s and $p$s. 

Thus  the question whether the accelerations are determined by the positions and velocities is translated to the question of whether the momenta $p_\alpha = \frac{\pp  \La}{\pp  \dot q^\alpha}$, are invertible, as functions of the velocities. If $M_{\alpha \beta}=\frac{\pp  p_\alpha}{\pp  \dot q^\beta}$ is not invertible, since $\alpha$ and $\beta$ run over the same indices (i.e. $M$ is a square matrix),  the map from the $\dot q$s to the $p$s, at fixed $q$, is many-to-one. 
That is:  there are constraints among the $p_\beta$ as functions of the velocities. Dropping reference to the velocities, i.e. writing the constraints as functions on phase space, we therefore write equations such as 
\be\label{eq:ctraints}
\Phi^I(q^\alpha, p_\beta)=0,
\ee
where $I$ parametrizes the  constraints.\footnote{Here we assume that the rank of $M_{\alpha\beta}$ is constant, and that the constraints obey regularity conditions. See \cite[Ch. 1.1.2]{HenneauxTeitelboim}.} These conditions are usually called {\em primary constraints}, to emphasize that no equation of motion was used in their derivation. If the constraints are conserved by the equations of motion, then  they correspond, by (the converse of) Noether's theorem, to symmetries of the system: as we will explain in Section \ref{sec:Ham_symp}. 

Now, the constraint surfaces are submanifolds of $T^*\mathcal{Q}$, and it thus follows from \eqref{eq:ctraints} that the inverse transformation, from the momenta $p$ to the velocities $\dot q$, must be  multi-valued, since the dimension of $T\mathcal{Q}$ is the same as that of $T^*\mathcal{Q}$ (viz. $2N$, and so greater than the dimension of  the constraint surfaces).  Thus the inverse image in $T\mathcal{Q}$ of the constraint surface \eqref{eq:ctraints} forms a submanifold.\footnote{The gradients of $\Phi^I$ in the momentum directions along the surface, i.e. the vectors $\frac{\pp \Phi^I}{\pp p_\alpha}$, provide the complete set of null vectors of $M_{\alpha\beta}$, since by \ref{eq:ctraints}
 $$0=\frac{\pp \Phi^I}{\pp \dot q^\beta}=\frac{\pp \Phi^I}{\pp p_\alpha}\frac{\pp p_\alpha}{\pp \dot q^\beta}=\frac{\pp \Phi^I}{\pp p_\alpha}M_{\alpha\beta}.$$
In Section \ref{sec:Ham_symp} we will give a coordinate-free version of these statements.\label{ftnt:M}}   Similarly, if we consider the intersection of the constraint surfaces \eqref{eq:ctraints}, as  $I$ varies (which is often called `{\em the} constraint surface'); and so, the inverse image of that intersection.    

 Correspondingly, within $T\mathcal{Q}$: a non-trivial kernel 
for the matrix $M_{\alpha\beta}$ in \eqref{equ:accel_lagrange} implies that the extrema of the Lagrangian are not isolated: there are 1-parameter families of curves in configuration space that extremize the action functional. In other words, as discussed above: there is indeterminism at the level of the $2N$ variables, $q^\alpha(t), \dot q^\alpha(t)$: (again, compare \cite{Wallace_LagSym}).

 Here is an \textbf{Example} from \citet[Sec. 1.1.1]{HenneauxTeitelboim}, which we will build on in Section \ref{sec:toy}. Let $(q^1, q^2)$ be coordinates for $\mathcal{Q}$, and consider $\mathcal{L}=\frac12(\dot q^1-\dot q^2)^2$. The momenta are $p_1=\dot q^1-\dot q^2$ and $p_2=\dot q^2-\dot q^1$, as is easy to verify. Thus we find the rather simple constraint: $\Phi= p_1+p_2=0$. The Legendre transform maps all of $T\mathcal{Q}$  into this constraint surface in $T^*\mathcal{Q}$. Moreover, the entire line $\dot q^2=\dot q^1+c$ is mapped to $p_1=-c=-p_2$. This transformation is therefore neither one-to-one nor onto. 

To render this transformation invertible, we will need to introduce \emph{Lagrange multipliers}, in Section  \ref{sec:Ham_symp}, which can be thought of as coordinates on the manifolds that are the inverse values in $T\mathcal{Q}$ (for \eqref{eq:Leg}) of a given point on the constraint surface lying in $T^*\mathcal{Q}$.\footnote{It is also relatively easy to show that two velocities that lie in the pre-image of the same momentum are related by a linear combination of the null vectors of $M_{\alpha\beta}$, namely $\frac{\pp \Phi^I}{\pp p_\alpha}$ (cf. footnote \ref{ftnt:M}); the coefficients of this linear combination are the Lagrange multipliers. In Section \ref{sec:Ham_symp} we will see a coordinate-free version of these statements. \label{ftnt:symp_flow} }

\subsection{The Hamiltonian formalism and constraints}\label{sec:Ham_symp}

To understand the basic features of constraints and the symmetries they generate in a manner that will be helpful in what follows, it pays to introduce the symplectic formalism for Hamiltonian mechanics. 

The great advantage of the symplectic formalism is that it treats momentum and configuration variables on a par. By so doing we see phase space $\mathcal{P}$ as a high-dimensional manifold---infinite-dimensional, in field-theory---endowed with a certain geometric structure. 
In the simple example above, $\mathcal{P}$ would be the $2N$-dimensional manifold  whose geometric structure is a symplectic 2-form, given, in  the global coordinates $(q^\alpha , p_\alpha)$, by:
\be \omega=\sum_\alpha \d q^\alpha\wedge\d p_\alpha.
\ee
Though we have given $\omega$ in the specific choice $(q,p)$ of coordinates,\footnote{These are always available locally, thanks to Darboux's theorem: which states, in modern geometric terms, that a manifold equipped with a symplectic 2-form is locally a cotangent bundle. Cf. e.g.  \cite[p. 230-232]{arnold1989}; or for a philosophical introduction,  \cite[ Section 6.6]{Butterfield2006}.} it is a coordinate-independent, differential geometric object on $\mathcal{P}$. 

The role of the symplectic form $\omega$ is to convert a vector field into a one-form, or vice-versa. And  since a scalar function defines a one-form, viz. its gradient, $\omega$ converts a scalar function like the Hamiltonian into a vector field, whose integral curves are a flow in phase space. We construe these curves as the dynamical trajectories of the system;  (i.e. assuming that the given scalar function encodes the forces operative on and in the system). 
Thus we take a Hamiltonian function $H:\mathcal{P}\rightarrow \bb R$ to specify the dynamics by assigning to each dynamical state its total energy.   For from any such smooth scalar function, we can obtain a one-form $\d H$, and then   use $\omega$ to define a vector field $X_H$, by: 
\be\label{eq:fund_symp} \omega(X_H, \bullet)=\d H (\bullet).
\ee 
This vector field on phase space   is then integrated to yield  a dynamical trajectory through each point. Compare Figure 1. 

\begin{figure}[h!]\label{fig1}
\center
\includegraphics[width=0.5\textwidth]{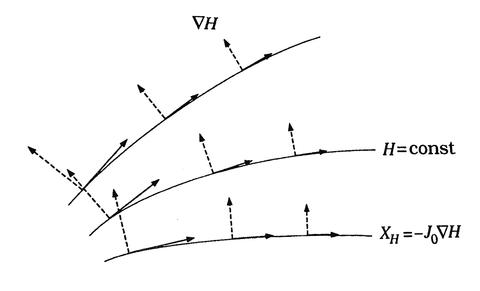}
\caption{An illustration of the relation between a scalar function, its gradient,  and its associated symplectic vector field, using the canonical symplectic form and the Euclidean metric  to identify $\d H$ with $\nabla H$, as a vector orthogonal to the level surfaces $H=$const. Supposing, in the $(q, p)$-coordinates, that $\nabla H=(\pp_q H, \pp _p H)$, then, writing the  inversion map as $J_0$ , we have: $X_H=-J_0(\pp_q H, \pp _p H)=(-\pp_p H, \pp_q H)$, which is orthogonal to $\nabla H$ and thus along $H=$ const.}
\end{figure}

Indeed, in the simple mechanical case without  constraints, we can plug coordinates into this equation to recover a local description of the dynamics, i.e. the familiar form of  Hamilton’s equations. In particular, the relation between Poisson brackets,   defined as usual by
\be\label{eq:Pois}
\{f, h\}:= \frac{\pp f}{\pp q^\alpha}\frac{\pp h}{\pp p_\alpha} - \frac{\pp f}{\pp p_\alpha}\frac{\pp h}{\pp q^\alpha}; \,\; \mbox{so that Hamilton’s equations with $h$ are:}  \,\; 
\frac{df}{dt} = \{f, h\} \, ,
\ee 
and the symplectic form $\omega$ is: 
\be\label{PoisOmeg}
 \{f, h\} = \d f (X_h) = \omega(X_f, X_h),
\ee
for $f, h\in C^\infty(\mathcal{P})$.\footnote{Here, as usual,  $\d f(X)$ is the contraction between 1-forms and vectors; and $\d f(X)$ is equal to $X(f)$ i.e. the directional derivative of a scalar function $f$ along $X$.}

For consistency, since the exterior derivative squares to zero i.e. $\d ^2\equiv0$,  $\omega$ must be closed, i.e. $\d \omega=0$. Moreover, if we would like the dynamical trajectory associated to $H$  to be unique,   $\omega$ must be  non-degenerate. That is: $\omega(v, \bullet)$ must be injective, i.e. have only the zero vector in its kernel. 

Although this last condition is always taken to hold on the full phase space $\mathcal{P}$, it needs to be relaxed in gauge theories, precisely because of  constraints. So although we require $\omega$ to be non-degenerate in $\mathcal{P}$, it does not need to be non-degenerate once we restrict it  to {\em the} constraint surface---meaning now
 the intersection of the constraint surfaces \eqref{eq:ctraints}, as  $I$ varies:
\be\label{eq:interGam}
\Gamma:=\{z\in \mathcal{P}, |\, \Phi^I(z)=0,\quad \text{for all}\,\, I\}.
\ee

 As depicted in Figure 1: the $\Phi^I$,  as scalar functions on phase space (for each $I$), have a (differential geometric) gradient, $\d \Phi^I$, which are in one-one correspondence with  vector fields $X_{\Phi^I} =: X_{I}$ due to the symplectic structure of phase space: namely, through  $\omega(X_I, \bullet)=\d \Phi^I$. The key idea is that, just as the flow specified by the Hamiltonian function conserves energy, these vector fields associated to $\Phi^I$ are tangential to, and so preserve, the intersection of all  the constraint- and energy-surfaces.  That is, in a less geometric (and maybe more familiar) language: they not only commute with the Hamiltonian and conserve energy, but also conserve the charges associated with the constraints.
 
  Here we will only consider  a certain type of constraints, called {\em first-class constraints}. These are defined as constraints whose Poisson bracket with every constraint vanishes on the constraint surface $\Gamma$ (though perhaps not elsewhere). Then the assumption that all the constraints are first-class implies that the flow of each vector field $X_I$, associated to each constraint, is tangent to $\Gamma$. This assumption also   
 means that we need not  concern ourselves with the several steps involved in the {\em Dirac algorithm}.\footnote{\label{algor}{Without the assumption of first-class, we still have an algorithm for finding whether the constraints generate symmetries. This algorithm can be summarised in the more familiar Poisson bracket notation, as follows. Suppose we are given some initial set of constraints $\Gamma_0:=\{z\in \mathcal{P}, |\, \Phi^{I_0}(z)=0,\quad \text{for all}\,\, I_0\}$, such that, e.g. for some $I_0, J_0$, we have  $\{\Phi^{I_0}, \Phi^{J_0}\}_{|\Gamma_0}=\Phi^{K_1}\neq 0$ where clearly $\Phi^{K_1}$ is not included among the original constraints, since it does not vanish on $\Gamma^0$. (Here the restriction to $\Gamma_0$ serves to emphasize that the vector fields need not commute everywhere on phase space, but only on the surface where the constraints vanish.) We would then  add this new constraint to the others, to form a new `surface', $\Gamma_1$, and repeat the test above, until, eventually we get:
$$
 \{ \Phi^{I_n}, H\}_{|\Gamma^n}\equiv\omega(X_{I_n}, X_H)_{|\Gamma^n}\equiv X_{H}(\Phi^{I_n})_{|\Gamma^n}=0,\quad \text{for all}\,\, I_n,
$$
and 
$$
\{\Phi^{I_n}, \Phi^{J_n}\}_{|\Gamma_n}\equiv \omega(X_{I_n}, X_{J_n})_{|\Gamma_n}=0,\quad \text{for all}\,\, I_n,J_n. 
$$
(It is possible that the iteration yields only the empty set, in which case the system is dynamically inconsistent).}}  In general, these steps are necessary because the  primary constraints \ref{eq:ctraints} that  emerge from the Legendre transform might fail to be preserved by either the Hamiltonian or by other constraints---but this will be ensured if the constraints are first-class.  Thus in general, one must then seek a type of reflective equilibrium: successively imposing further restrictions to  submanifolds of phase space, until the corresponding symplectic flows of all the constraints preserve the Hamiltonian and become tangent to the constraint submanifold. Compare footnote \ref{algor}, and  \cite{Pons2005}; and for a complete account, \cite[Chapter 2]{HenneauxTeitelboim}. 
 
 In the language of symplectic geometry, we define the embedding $\iota:\Gamma\rightarrow \mathcal{P}$ and require that  the pullback of the symplectic form $ \iota^*\omega$ obey:
\be\label{eq:Ham_Gamma}
 \iota^*\omega(X_H, \bullet)=\d (H|_{\Gamma}), \quad \iota^*\omega(X_I, \bullet)|_{\Gamma}=\d (\Phi_I{}|_{\Gamma})\equiv 0
\ee
(since $\Phi_I{}|_{\Gamma}\equiv 0$). Thus, on the constraint surface, because we assumed constraints to be first-class, the vector fields $X_I$ generated by the constraints are null directions of the symplectic form restricted to the constraint surface. This generalizes, in a  coordinate-free formalism, the content of footnote \ref{ftnt:symp_flow}. 

These directions are \emph{gauge}. 
The set of points of phase space that are reached by the $X_I$'s integral curves, from a given point, is called a \emph{gauge orbit}.  And the elements within each such orbit   are taken to be physically equivalent. Here, physical equivalence of two points of phase space is understood as a matter of any physical quantity, taken as a phase function, i.e. a real-valued function on phase space, having the same value for the two points. That is: a physical quantity must be {\em gauge-invariant}: taken as a phase function, it  must be constant on each gauge orbit.\footnote{\label{fnMagic} {Indeed, the null directions of  $i^*(\omega)$ are necessary and sufficient to characterise the generators of gauge symmetry. For suppose that what we know is that  a certain class of vector fields $X_I$ is such that $\omega(X_I, \bullet)=0$.  Since  the exterior derivative $d$ commutes with pullbacks, if $\omega$ is closed, $i^*\omega=:\tilde \omega$ is also closed. Thus using the Cartan Magic formula relating Lie derivatives, contractions $i$ and the exterior derivative $\d $:   
$$
 L_{X_I}\tilde\omega=(\d i_{X_I}+ i_{X_I}\d)\tilde\omega=0; 
$$
i.e. the first term also vanishes because $\tilde\omega(X_I,\bullet)=0$. So $\tilde\omega$ itself is invariant along $X_I$. Moreover, if we take the commutator of $X_I, X_J$, i.e. $[ X_I, X_J]= L_{X_I}X_J$,  contract it with $\tilde\omega$, and remember the formula:
$$L_{X_I}(\tilde\omega(X_J, \bullet))=\tilde\omega( L_{X_I}X_J, \bullet)+  (L_{X_I}\tilde\omega)(X_J, \bullet) \, ,
$$
we obtain that, since both $L_{X_I}(\tilde\omega(X_J, \bullet))=0$  and $L_{X_I}\tilde\omega=0$, it is also the case that $\tilde\omega( [{X_I}, X_J], \bullet)=0$. Thus, by the Frobenius theorem  the kernel of the pullback  $i^*(\omega)$ forms an integrable distribution which  integrates to give  the orbits of the symmetry transformation.  This means, for the discussion below (in Section \ref{sec:off}) of symplectic reduction, in which a Lie group $G$ acts {\em ab initio} on the phase space, that:} we can define a projection operator $\pi:\Gamma\rightarrow \Gamma/G$; and, ultimately the degeneracy of $i^*\omega$ allows one to define a \emph{reduced symplectic form}, $\bar\omega$,  on the space of orbits, given by $\pi^*\bar\omega=i^*\omega$. See \cite[Ch. 1]{Marsden2007}.}

 For simplicity, we have here suppressed  a few important qualifications (that are widely recognized). Firstly: in some systems, the requirement that the primary constraints \eqref{eq:ctraints} be preserved in time implies a new relation between the $q$s and $p$s, independent of these constraints. Such a relation is called a {\em secondary} constraint. Secondary constraints can be first-class: an important example being the Hamiltonian i.e. scalar constraint  (governing time-evolution) in canonical general relativity. And some such secondary first-class constraints are {\em not} gauge generators. But though important, these points do not affect this paper. For details, compare \cite[Sections 1.1.5, 1.2, 1.6.3 and 3.3]{HenneauxTeitelboim} and \cite{Pitts_Ham}.   A second simplifying assumption---which applies to electromagnetism, and indeed to most familiar physical theories---is that the commutation algebra of the constraints closes irrespectively of the satisfaction of the equations of motion: it forms what is usually called a \emph{closed algebra}. Thirdly, we   have also assumed that the constraints are {\em irreducible}, i.e. that all the constraint equations \ref{eq:ctraints} (for both primary and secondary constraints) are independent of each other: (so that roughly speaking, there are no ``symmetries among the symmetries''  that they generate). And lastly, we   have also assumed that the commutation algebra of the constraints forms a true Lie algebra, i.e. the structure `constants' are true constants, not functions on phase space; (this assumption fails for general relativity).\footnote{\label{Pitts}{We should also note a controversy. \cite{Pitts_electro} claims that, even under these assumptions, and for electromagnetism, the main (and orthodox) idea above---that points in the same gauge orbit, are physically equivalent---fails. He claims that even a first-class constraint can fail to be a gauge generator, i.e. it can generate instead what he calls a `bad physical change'. Our own view is that the main idea holds good. (The dispute turns on the transformation properties of Lagrange multipliers in the canonical Lagrangian; and we think the treatment by e.g. \cite[equations 19.11 and 19.13, as clarified and supported by 3.26 and 3.31]{HenneauxTeitelboim}  answers Pitts’ arguments.)}}

 The existence of null vector fields implies the Hamiltonian flow is not unique: if $X_H$ solves \eqref{eq:Ham_Gamma}, then so does $X_H+a^I X_I$ for any set of  coefficients $a^I$ that are arbitrary functions of time. So, since the dynamics preserves the constraint surface, instead of taking $H$ as the Hamiltonian function generating the dynamics, we may equivalently take the \emph{total Hamiltonian}:
\be\label{eq:Ham_tot}
H_T=H+a^I\Phi_I.
\ee
Along the constraint surface the dynamics according to $H_T$  will be indistinguishable from that according to $H$.  Here, by `indistinguishable',  we mean that  two trajectories within $\Gamma$ that start at a common point in $\Gamma$ and that are determined, respectively, by the two choices of Hamiltonian, will at any later (indeed: any earlier!) time, lie in the same gauge orbit as each other---and so will at all times agree on the values of all gauge-invariant quantities.

In sum,  one of the  significant (as well as practical) differences between the Hamiltonian and the Lagrangian   frameworks is that  in the Hamiltonian framework the symmetries are not `guessed'  from the form of the action functional. Instead, they are obtained from the constraints that emerge 
when  the Legendre transformation is applied. That is, the constraints that emerge from the Legendre transformation are associated to vector fields; and, with a few auxiliary assumptions,   the flow of   each of these vector fields conserves the constraints and the Hamiltonian, and   the vector field is thus taken to be the generator of symmetries on ${\Gamma}$. The action of the symmetry on any quantity $Q$ is given by  
\be\label{symquant}
 X_I(Q)=\d Q(X_I)=\omega(X_I, X_Q)=\{\Phi^I, Q\}.
\ee

 Finally let us sum up this discussion of the Lagrangian and Hamiltonian frameworks’ treatments of constraints, by stressing a concordance between them: a concordance despite the many-one mapping from the first to the second, i.e. the fact that at fixed $q$, many $\dot q$ map to a single $p$. 
 Namely: there is a  one-one correspondence between the gauge orbits in the Hamiltonian framework and  degeneracy directions of the Lagrangian in the configuration space; (cf. Sections 3.1-2 in \cite{HenneauxTeitelboim}).

\subsubsection{Off the constraint surface}\label{sec:off}
 Our sketch of the Hamiltonian treatment of constraints would be incomplete without some mention of {\em symplectic reduction}: a large and important topic  (briefly mentioned in footnote \ref{fnMagic}).  However, it is usually pursued, not  (as in this paper) by using the the Legendre transformation being many-one to motivate  restricting one’s attention to the constraint surface $\Gamma$ in phase space (as in this paper); but by postulating {\em ab initio} a smooth action of a Lie group $G$ on phase space, and studying the consequences. So the relation to the Lagrangian framework, and to constraints originating from $\frac{\partial p_\beta}{\partial {\dot q}^\alpha}$ being not invertible, tends to be obscured. But the rest of this paper will not need an account of symplectic reduction: for which, cf. e.g. \cite{Marsden2007} for a  complete, but concise, exposition, \cite{Butterfield_symp} for a philosophical introduction, and \cite{GomesButterfield_glimpse} for the relation to the Lagrangian framework.

However, symplectic reduction prompts a philosophical topic we want to address. It is about possibility, i.e. about how we should think of the non-actual i.e. unrealized states in the state-space. We have hitherto said nothing about this, since our discussion has  prompted no alteration from how one normally thinks of possibility within the Hamiltonian dynamics of an {\em unconstrained} system. There, one naturally regards which energy hypersurface the system is actually on (or equally: the actual value of any first integral of the motion) as a  matter of initial conditions, mere happenstance. And for all we have said so far, it seems that in general, this attitude applies equally to constrained systems. For in general, it seems that the constraints $\Phi^I$ could have taken values other than zero  (cf. equation \ref{eq:ctraints}): the state could have been {\em off} the actual constraint surface $\Gamma$. But the theory of symplectic reduction reveals a wrinkle: indeed, two wrinkles. \cite[Section 11]{Belot2003}  discusses them, as do  \cite{GomesButterfield_glimpse}; and we sketch them here.

First, in some cases there is reason to deny that the states lying off the constraint surface $\Gamma$ are genuinely possible. 
Belot's (and our) simplest  example is {\em relationism}, {\em a la} Leibniz and Mach, about space. The case can be made for a system of $N$ point-particles in Euclidean space. For  this system, while the `absolutist' will take the configuration space $\cal Q$ to consist of all the ways $N$ particles can be placed in $\R^3$, i.e. to be $\R^{3N}$, the relationist will say that two such placements that differ by a spatial translation and-or a rotation should be regarded as one and the same. That is, the relationist advocates a {\em relative configuration space}, whose points are sets of relative distances between the particles. This space can be presented as the quotient of  $\R^{3N}$ by the obvious action of the Euclidean group; (modulo some technicalities about excising unsuitably symmetric points of $\R^{3N}$). And when one works through the details of the constraint formalism, it turns out that on the relationist's view, only states on the relevant constraint surface $\Gamma$ within $T^*{\cal Q}$ are genuinely possible. 


Of course, not everyone is a relationist! But also in other cases, there is a similar rationale to `endorse the dynamics intrinsic to $\Gamma$', and reject the states not in $\Gamma$. Thus Belot points out that in some field theory cases, theorizing about states off the constraint surface corresponds to treating charges that source the field in question without being affected by it, i.e. treating external sources. For example, in our case-study of electromagnetism: for non-vacuum, the Gauss constraint becomes div$(E) = \rho$, where $\rho$ is treated as sourcing, but as unaffected by, the electromagnetic field. (Of course, there is a close analogy with Poisson’s equation $\nabla \phi = \rho$ in Newtonian gravity, and its modern descendant, Newton-Cartan gravitation: the mass density $\rho$ sources the potential $\phi$ but does not self-gravitate.) Since this is an idealization, one has reason to reject the states off the constraint surface as not genuinely possible, and to endorse the dynamics intrinsic to $\Gamma$---like the relationist above. 

Besides, as Belot goes on to say: faced with this idealization, one should seek theories in which the coupling is ``two-way’’. Indeed, there are such theories (references in his footnote 74); and---what matters for our present topic---in these theories, one again  gets \emph{only one constraint surface}, like in the case of vacuum electromagnetism. That is:  once one augments the phase space so as to describe the charges (including: augmenting the Hamiltonian to describe the two-way coupling), one gets just one constraint surface 
in a higher-dimensional space---{\em not} a family of surfaces in the  original lower-dimensional phase space, indexed by the charge distributions. 
  So again, one has reason to endorse the dynamics intrinsic to the constraint surface.  And again: our main theme about gauge structure is illustrated: viz. that null vector fields on the constraint surface are infinitesimal generators of gauge transformations.

Let us summarise this discussion by quoting Belot. He writes (p. 215):---
\begin{quote}

This [i.e. treating charges as external, i.e. as sourcing the field in question yet without being affected by it] amounts to working off of the constraint surface in order to study the field dynamics in the presence of external sources painted onto spacetime independently
of the behavior of the fields. This is, of course, an ad hoc maneuver—if one wants to
study Yang–Mills with sources honestly, one must introduce matter which not only
acts upon the field but is also acted upon by it. And when one pursues this upright
course, one ends up with a constraint which is a direct analog of the usual Gauss
constraint—the null directions of the constraint surface correspond to the infinitesimal generators of gauge transformations. Under this more fundamental
approach, there is no physical interpretation for points lying off of the constraint
surface—and so we have an excellent reason to prefer an intrinsic reading of the
theory.
\end{quote}

\section{The case of electromagnetism}\label{sec:HamEM}

 So much by way of reviewing Hamiltonian constrained dynamics. We turn to our case-study, classical electromagnetism.
There are many accounts of its symplectic structure in the literature, both physical and philosophical: some of them of course excellent.\footnote{Excellent physics expositions include \cite{JacksonBook}. Excellent philosophical discussions include: \cite{Belot1998, Belot2003, Healey_book}. We particularly recommend \cite{Belot2003} for the Hamiltonian  formulation of vacuum Yang-Mills theories.} But there is a ``core'' of ideas and results, 
that is relatively easy to expound and illuminating, without  having to plumb the depths of the (elegant) symplectic geometry of the theory.  This core is, so far as we know, not articulated in the literature: certainly, it is not stressed. 

So as we announced at the start of Section \ref{intro}, the technical aim of this paper is to expound this core, and show that it sheds light on various ideas, both formal and interpretative. More specifically, we will try to shed light on (1) classical non-locality  and (2) the preferred splits of degrees of freedom. (These correspond to the three morals   at the end of Section \ref{intro}.) 

 In Section \ref{sec:toy} we will illustrate the ideas that are used in (2) in a simple toy example. We will see how certain considerations of convenience and simplicity can go a long way to selecting gauge-fixing conditions.  
  In Section \ref{sec:symp_EM} we will apply those ideas to electromagnetism, where they give rise to the Gauss constraint and gauge transformations. In Section \ref{sec:Gauss} we will interpret the Gauss constraint as encoding a type of non-locality, {\em a la} (1). It defines a part of the electric field, viz. the Coulombic field, that is determined by the instantaneous distribution of charges. (Adopting \cite{Maudlin_ontology}'s terms for a moment: it is not `fundamental ontology', but `derivative ontology', since derived from the charge distribution.) There is here a strong analogy with the elementary Newtonian gravitational potential $\phi$, 
 which is sourced by the mass distribution via Poisson's equation: and which is often said to be ``not physically real'', or  ``a convenient fiction'', since it has no energy or momentum, but only encodes, via its gradient $\nabla \phi$ the infinite battery of counterfactual conditionals about how test-masses located at the spatial point in question would accelerate. Similarly here: the Coulombic field encodes infinitely many counterfactual conditionals about how test-charges would move---if there were also no other part of the field, i.e. no radiative part, enjoying its own dynamics. 
 
   Using the decomposition of fields implied by this understanding, we will in Section  \ref{sec:radcoul} use the symplectic structure of the theory to find the ``block-diagonal conjugate structure''  of the electric and gauge potential fields, and show how this selects the Coulomb gauge.

\subsection{A toy example of natural coordinate choices on phase space}\label{sec:toy}
To make our aims more vivid, we return to the simple example we gave at the end of Section \ref{sec:Lag}.  Consider two identical free particles of mass $m$ on a line, with coordinates $q_1$ and $q_2$.   So the phase space $\cal P$ is 4-dimensional, with  coordinates  $(q_1, q_2, p_1, p_2)$.  The canonical Hamiltonian for the system is
\begin{equation}
 H = \frac{1}{2m}(p_1^2 +p_2^2).
\end{equation}
Now we add a constraint to the system, namely:
\begin{equation}
\Phi = p_1 + p_2 = 0 \, ;
\end{equation}
(which is clearly first-class). So the constraint $\Phi=0$ requires the total canonical momentum $P:=p_1 + p_2$ to vanish,   and defines a 3-dimensional constraint surface $\Gamma$ in $\cal P$. The total Hamiltonian is ${H}_T  = H + a \, \Phi $, where $a$ is an arbitrary function of time. It generates the time evolution
\begin{equation}\label{eq:mom_ex}
\dot q_1 = \frac{p_1}{m} + a 
\,, \qquad 
\dot q_2 = \frac{p_2}{m} + a \,,
 \qquad 
\dot p_1 = \dot p_2 =0 \,.
\end{equation}

Here we started with the variables that initially seemed natural, centered on each of the particles. But a more natural choice for coordinatizing momentum space  (i.e. at each fixed value of $(q_1, q_2)$)  would be $P_+(p_1,p_2):=p_1+p_2$ and $P_-(p_1, p_2)=p_1-p_2$. Now   at each  value of $(q_1, q_2)$, the constraint surface is given by: 
\be 
\{(P_+, P_-)\, |\, \Phi(z)=0\}=\{(0, P_-) \, |\, P_-\in \bb R\} \, .
\ee

In the $q$-coordinates, the natural conjugate variables to $P_-$ and $P_+$ are $Q_-=q_1-q_2$ and $Q_+=q_1+q_2$, respectively. The first, $Q_-$, is the relative distance between the two particles. It  is gauge-invariant, $\{ q_1 - q_2 , \Phi \} = 0$; and its equation of motion contain no arbitrariness,
\begin{equation}
\label{eq:dotQ}
\dot q_1 - \dot q_2 =\dot Q_-= \frac{p_1}{m} -
 \frac{p_2}{m}  =\frac{P_-}{m}\,.
\end{equation}
On the other hand,  from \eqref{eq:mom_ex},
\be\label{eq:gauge_q} \dot Q_+=m \, \dot q_1
+ m \, \dot q_2 = 2m a \ee
is ``pure gauge'', since $a$ is an arbitrary function of time. Note that \eqref{eq:dotQ} and \eqref{eq:gauge_q} are given on the full phase space: of course, the symplectic form on the constraint surface would be degenerate, since $P_+=0$ there.   To sum up: $Q_-$, $Q_+$ and $P_-$ are natural coordinates to parametrize the constraint surface.  

In what follows, we will try to provide a similarly natural decomposition of electromagnetism's gauge potential and electric field. $\Phi$ will be the Gauss constraint, which we will take to be naturally parametrized by a Coulomb potential. So this potential  will play the role of $P_+$, and thus the radiative degrees of freedom of the electric field will play the role of $P_-$. As regards the configuration variables: $Q_-$ will be given by a more complicated function of the original configuration variables, which projects it into Coulomb gauge, but it will be likewise gauge-invariant. And finally, $Q_+$ will be the pure gauge part of the gauge potential. 

\subsection{Hamiltonian treatment of electromagnetism}\label{sec:symp_EM}

The Maxwell equations are written, in terms of the electromagnetic field tensor, $F_{ab}$, in four-dimensional Minkowski spacetime $M$, as:
\be\label{eq:EM}
\partial^a F_{ab}=j_b, \quad \text{and}\quad \partial_{[a}F_{cd]}=0; 
\ee
where $j$ is the electric 4-current and square brackets denote anti-symmetrization of abstract spacetime indices ($a, b,$ etc.). 

The second equation of \eqref{eq:EM} is called `the Bianchi identity', and it is read as a  constraint on the field tensor. A geometric explanation for this constraint is that $F_{ab}=\partial_{[a}A_{b]}$, where the 1-form $A_a$ is called \textit{the gauge potential}. That is: $F_{ab}=\partial_{[a}A_{b]}$ implies the second equation of \eqref{eq:EM}   holds identically, thanks to the commutation of partial derivatives.  By employing the gauge potential we can drop the Bianchi identity, and that is what we will do. 

Then the equations of motion \eqref{eq:EM} are written as: 
\be\label{eq:eom_A}
 \partial^a\partial_a A_b-\partial^a\partial_b A_a=j_b.
 \ee
 These equations (together with the equations for the dynamics of the charges constituting the currents) are obtained from the action functional:
  \be\label{eq:action_A}S[A]:=\int \d^4 x\left( \partial_{[a}A_{b]}\partial^{[a}A^{b]} +A^a j_a +\mathcal{L}_{\text{\tiny{matter}}}\right),  
 \ee
 where $\mathcal{L}_{\text\tiny{matter}}$ is the Lagrangian density for the matter fields; (for illustration, one can take this  as the Klein-Gordon Lagrangian or as the Lagrangian for a charged point particle).
 
Now we choose a spacetime split into spatial and time directions,  $M=\Sigma\times \bb R$.  (We recall Section \ref{intro}'s admission that this carries the price of losing manifest Lorentz invariance.) We also assume  that the fields have appropriate fall-off conditions at spatial  infinity.

Upon such a spacetime decomposition, the components of the electromagnetic tensor recover the familiar electric and magnetic fields: $F_{i0}=E_i$, and $F_{ij}\epsilon_i^{jk}=B_i$ and $j_0=\rho$  (where we used the three-dimensional totally-antisymmetric tensor, $\epsilon$, or the spatial Hodge star, to obtain a 1-form, and $i, j, k$ are spatial indices, i.e. in $\Sigma$), and the first equation of \eqref{eq:EM} becomes the familiar Maxwell equations.\footnote{  For $\mathcal{L}_{\text{\tiny{matter}}}=mv^a v_a$, for $v^a=(\dot \gamma)^a$ the 4-velocity of a charged particle whose trajectory is $\gamma$, i.e. such that $j^a=e v^a$, we obtain the Lorentz force equation  as the equation of motion for the particle.}

Now we perform a Legendre transform. Then the spatial vector fields $\bf A$ and  $\bf E$ are canonically conjugate. In fact, $\bf A$ is the configurational variable, and $\bf E$ is the momentum; since ${\bf E} = {\pp {\cal L}}/{\pp {\dot{\bf A}}}$. The Poisson bracket is defined, for $F, G$ two functionals of the fields $A$ and $E$; 
\be\label{PoisField}
\{F,G\} := \int d^3y \; (\frac{\delta F}{\delta A_i(y)}\frac{\delta G}{\delta E^i(y)} \, - \, \frac{\delta F}{\delta E^i(y)}\frac{\delta G}{\delta A_i(y)}) \, .
\ee

Now, as described in Section  \ref{sec:Lag} and Section \ref{sec:Ham_symp}, the Lagrangian has symmetries, which translate into constraints in the Hamiltonian formalism. Analogously to \eqref{eq:Ham_tot}, we obtain a total Hamiltonian, written as $H_T=H(A, E, j, \lambda)+{H}_{\text{\tiny{matter}}}$, with: 
\be\label{eq:Ham}
H(A, E, j, \lambda)=\int \d^3 x \left(\|\mathbf E\|^2+\|\mathbf{B}^2\|+\lambda(\mathrm{div}(\mathbf E)-\rho) +A^i j_i\right), 
\ee
where $\partial_{[i}A_{j]}\partial^{[i}A^{j]}=:\|\mathbf{B}^2\|$,  $\mathrm{div}(\mathbf E):=\pp^i E_i$. 

The part of \eqref{eq:Ham} that we want to draw attention to is the term $\lambda(\mathrm{div}(\mathbf E)-\rho) $. For $\lambda$ is a scalar function on the spatial surface $\Sigma$: $\lambda$ is the \emph{Lagrange multiplier} we first encountered in Section \ref{sec:Lag}. Indeed it is just the $a^I$ of \eqref{eq:Ham_tot}, but  now in the field-theoretic context, when $I$ becomes a continuous index. The constraints corresponding to $\Phi^I$ are:
\be 
\label{eq:Gauss}
\mathrm{G}(x):=\mathrm{div}(\mathbf E)(x)-\rho(x)=0, \quad \text{for all}\,\,x\in \Sigma. 
\ee
Accordingly, \eqref{eq:Gauss} defines not a single constraint, but an infinite set of them: one per spatial point.    The values of $\lambda$ at the various points $x \in \Sigma$ thus give a particular linear combination of the constraints. Hence $\lambda$ is also called a `smearing' of the constraints; and it  is convenient to define the smeared Gauss constraint: $\mathrm{G}(\lambda):=\int_\Sigma(\mathrm{div}(\mathbf E)-\rho)\lambda$. 

To see that we are in the domain of the previous discussion, namely that the constraints are all first class, we can check that the constraints  commute, and also commute with the Hamiltonian constraint. This is easy to verify, since a given linear combination of   symplectic flows, which we call a \emph{smeared} symplectic flow,  $X_\mathrm{G(\lambda)}$, acts on the canonical variables as: 
\begin{align}
X_{G(\lambda)}({A}_i(x))&:=\{  {A}_i, \mathrm{G}(\lambda)\}:=\int \d^3 y\frac{\delta( \lambda(\mathrm{div}(\mathbf E)-\rho))}{\delta E^i(x)}=\int \d^3 y\lambda(y)\pp_i\delta(x,y)=-\pp_i\lambda(x);\label{eq:G_A}\\
 X_{G(\lambda)}({E}^i(x))&:=\{ {E}_i, \mathrm{G}(\lambda)\}:=-\int \d^3 y \frac{\delta( \lambda(\mathrm{div}(\mathbf E)-\rho))}{\delta A_i(x)}=0\label{eq:E_gt}
\end{align}
To obtain these results, note that in each line the third term, i.e. the integral of a functional derivative, gets just one term from  \ref{PoisField}; and that  the final equation in \eqref{eq:G_A} is obtained by integration by parts removing the derivative of a delta-function.

So we see that, as expected,  the constraints act on phase space as the familiar gauge transformations: they preserve the value of the electric field (which is gauge-invariant in electromagnetism) and change the value of the gauge potential by a gradient of a scalar function: $A_i\mapsto A_i-\pp_i \lambda$.\footnote{The Gauss constraint is Lie-algebra valued, even in the Abelian case; that is necessary for it to generate infinitesimal gauge transformations. It just so happens that here   the Lie group $G$ is $S^1 = U(1)$, and so the Lie algebra is $\mathfrak{g}\simeq \bb R$. In the non-Abelian theory,   the Gauss constraint is $G^\alpha(x)=\D_A E^\alpha-\rho^\alpha$, where, in a given basis, $\D_A E^\alpha= \pp^i E_i^\alpha-\epsilon^\alpha_{\beta\gamma}E_i^\beta A^{i\gamma}$, where $\alpha, \beta, \gamma$ are Lie-algebra indices and $\epsilon$ is the structure constant of $\mathfrak{g}$ in the given basis. Due to the appearance of $A$ in the Gauss constraint, it acts linearly on $\mathbf{E}$: which is thus only covariant, and not invariant as it is in the Abelian theory. }
Since the flow of the constraint does not change the Hamiltonian (and preserves the constraint as well), it generates a symmetry of the system.

From the more geometric viewpoint, the infinite-dimensional phase space whose canonical coordinates are the electric field and the gauge potential has a bona-fide (infinite-dimensional) symplectic geometry (e.g. modeled on Banach manifolds, cf. \cite[Chapter 2]{Lang_book}). Thus we have a symplectic form, which  in vacuum is written as:\footnote{It is easy to extend this to the presence of matter. For example, with a Klein-Gordon field, we would add: $\int \sqrt{g} \;\dd \bar\psi \curlywedge \gamma^0  \dd \psi$. See \cite[Section 3]{GomesRiello_new} for more details on the symplectic geometry of the infinite-dimensional space.}
\be\label{eq:symp_form}
\Omega  = \int \d^3x  \,\delta E^i \curlywedge \delta A_i.
\ee
Here we take $\delta E^i$ and  $\delta A_i$ to be the fundamental one-forms on phase space, and $\curlywedge $ to be their anti-symmetrization (the exterior product of   one-forms in an infinite-dimensional space). 

Just as  for finite dimension, $\frac{\pp}{\pp x^i}, \frac{\pp}{\pp p_i}$ are the vectors tangent to the $x^i$ and $p_i$-coordinates: so also $\frac{\delta}{\delta E^i(x)}$ and $\frac{\delta}{\delta A^i(x)}$, for each $x\in \Sigma$, are vectors in  the infinite-dimensional  phase space. And just as  for finite dimension, we can find new directions, or vector fields, by the linear sum, 
$\sum_i a^i \frac{\pp}{\pp x^i}$: here also, we can find new directions by linear sums, i.e. by integrating the fundamental directions smeared  with certain coefficients, e.g. 
\be\label{eq:infv} 
\bb v=\int \d ^3 y\, v^i(y)\,\frac{\delta}{\delta E^i(y)}.\footnote{ In electromagnetism, the standard basis, namely $(E_i(x), A^i(x))$, for $x\in \Sigma$,  is in this language just a basis that takes the coefficient functions to be all multiplied by Dirac deltas, e.g. a subset of smearings $v^i_x=E^i(x)$ defined as $v_x^i(y)=E^i \delta(x,y)$, for each $x\in \Sigma$ and $E^i\in \bb R$. This is what makes that basis completely local. Here we want to find a particular coordinate system, that is not completely local in this way. But in general, one could use symplectic duality to find the dual of any maximal Lagrangian submanifold in phase space.}
\ee (where we use a double-struck notation to indicate that these are vector fields on the infinite-dimensional phase space of classical electromagnetism).

Moreover, configuration space,  which is the space of smooth gauge potentials $\mathcal{A}=\{\mathbf{A} \in C^\infty(\Sigma, \R^3)\}$ admits an action of the (infinite-dimensional) group of gauge transformations, $\mathcal{G}=\{\lambda\in C^\infty(\Sigma)\}$, namely: $\mu:\mathcal{G}\times \mathcal{A}\rightarrow \mathcal{A};$ $\mu(\lambda, \mathbf{A})=\mathbf{A}-\mathrm{grad}(\lambda)$.\footnote{Due to the presence of stabilizers,  $\mathcal{A}$ is not an infinite-dimensional principal $\mathcal{G}$-bundle, i.e. $\mathcal{G}\hookrightarrow\mathcal{A}\rightarrow \mathcal{A}/\mathcal{G}$. But, at least in the Abelian case, this is easy to remedy by considering only the group of pointed gauge transformations (i.e. those that are the identity at a given point). But $\mathcal{A}$ always has a local product structure (i.e. it admits a slice, see \cite{kondracki1983} and references therein), even in the non-Abelian case.}  This structure exists also for the non-Abelian Yang-Mills theories. In any case, by using the isomorphism that $\Omega$ provides between $T\mathcal{A}$ and $T^*\mathcal{A}$, this action can be lifted to phase space in the usual manner (yielding \eqref{eq:E_gt}, or its appropriate non-Abelian generalization).


\subsection{Interpreting the Gauss constraint}\label{sec:Gauss}

We now combine Sections \ref{decide!}’s and \ref{sec:symp_EM}’s discussions of the Gauss constraint with a comment on a proposal of Maudlin’s. This will pave the way for our main result in Section \ref{sec:radcoul}.

As we said at the end of Section \ref{decide!}: the Gauss constraint (now in the form of \eqref{eq:Gauss}) involves a (non-signaling) kind of non-locality. For (by the elementary divergence theorem) the integral of $\bf E$ over a surface enclosing a spatial region determines the total  charge  inside the surface {\em at the given instant}.\footnote{Here, `determines’ can be read as `gives us knowledge of’. For one can imagine measuring $\bf E$ precisely throughout the surface, doing the integral, and inferring the total charge. This is, of course, the intuitive basis for quantum theory’s charge superselection rule. For using this procedure, one can measure the total charge at an arbitrarily large distance from the system; and this suggests that measuring charge is compatible with measuring any other quantity on the system, so that the charge operator  commutes with the operator representing that quantity.} Like the Aharonov-Bohm effect, this type of non-locality is classical; but unlike the Aharonov-Bohm effect, it does not require an underlying topologically non-trivial spatial domain for the probe systems. (Again, compare  Appendix \ref{app:hol}.) 

As mentioned in this Section’s preamble, this suggests isolating the part of the electric field that is determined by the instantaneous distribution of charges, and thinking of it as `derivative’ from this distribution. We will shortly pursue that idea. But first we notice that \cite[p.10]{Maudlin_ontology} proposes to `turn this around’. Thus he writes:
\begin{quote}
Let us propose that this equation [i.e. div$({\bf E}) = \rho$] represents not a physical law but an ontological
analysis: electric charges just are the divergences of electric fields. In this way we reduce both the
physical ontology and the nomology, and further gain an explanation of why electric charges cannot
exist without electric fields.
\end{quote}
So Maudlin proposes that the Gauss constraint be read as a definition of $\rho$, the electric charge. (He springboards this proposal from the corresponding one (p. 9) about $\bf B$: that div$({\bf B}) = 0$ is completely equivalent to the non-existence of magnetic charges.)

We submit that---perhaps unfortunately---this proposal does not work. There are various problems. The most obvious is that there are particles with the same electric charge but differing in other ways; (they have different masses, and-or different charges for other interactions). 
 That is: particles have characteristics independent of their   electric charge. (And if one attempted to define the source of each distinct field in a similar eliminative manner, one would then have to explain why all the different sources and sinks coincide in space and time.) Other problems include: (i) since electromagnetism is a linear  theory, must charged sources pass though each other? (ii) how do we explain interactions of the electromagnetic field with (apparent) matter, such as light reflection and refraction?\footnote{Notice incidentally that Maudlin is re-inventing the wheel. It is a creditable re-invention, since the wheel has a venerable design: but for all that, the wheel does not roll. That is, without metaphors: Maudlin’s proposal is the initial idea of the {\em electromagnetic world-view} that aimed to reduce mechanics (and all physics) to electromagnetism. In particular, it took the velocity-dependence of particles' masses to suggest that all mass might be of electromagnetic origin. It was advocated in the early twentieth-century by physicists such as Abraham and Mie; for an introduction cf. \cite[Chapter 8]{Kragh1999}. We should also note the ongoing---yet also very speculative---programme to reduce both matter and radiation to structures in spacetime, sometimes called `super-substantivalism': for which, \cite{MisnerWheeler1957} is a classic, and \cite{Lehmkuhl2018} is a fine philosophical introduction. 
 
Maudlin’s proposal also connects with the discussion at the end of Section \ref{sec:off}. Namely: {\em if} the proposal held good so that charge was indeed fully defined by the electric field, then our objection in that discussion, that treatments of charge as sourcing the field but as unaffected by it are an idealization---which should be replaced by a ``two-way’’ coupling---would fall by the wayside.   
}



But the failure of this proposal does not impugn the more modest idea above: that we should isolate the part of the electric field that is determined by the instantaneous  charge distribution, and think of it as `derivative' from the distributions.  So we think of this part---the Coulombic part---of the field as `the price to pay' for a local formulation of the theory; but it can be replaced by the more fundamental distribution of charges acting at a distance. In other words: there is a certain component of the interaction between charges that does {\em not} take into account a field that has its own dynamics: all this component needs is the present distribution of the charges themselves. The remaining part of the electric field is then interpreted  as `fundamental', in the sense that it has its own dynamics: a dynamics that is not reducible to the dynamics of other components of the theory.

The upshot will be that, since $\mathbf{E}$ and $\mathbf{A}$ are conjugate, and we take part of $\mathbf{A}$ to be `pure-gauge',  the above decomposition of $\mathbf{E}$ can be made to  correspond, in a particular way, to a certain decomposition of $\mathbf{A}$. We will do this in parallel to the toy case of \S \ref{sec:toy}, but here defining the gauge-complementary part of $\mathcal{A}$ to be symplectically orthogonal to a natural parametrization of the Coulombic part of $\mathbf{E}$, and by defining the Coulombic-complementary part of $\mathbf{E}$ to be symplectically orthogonal to the pure gauge part of $\mathbf{A}$. 

\subsection{Radiative and Coulombic parts of $\mathbf{E}$ and $\mathbf{A}$}\label{sec:radcoul}

Now we come at last to our application of {\em the Helmholtz theorem}. The theorem states that any vector field $\mathbf{Z}$ on $\bb R^3$ can be decomposed (`split') as:
\be \mathbf{Z}=\mathbf{X}+\mathbf{Y}
\ee 
for a unique pair of curl-free ($\mathbf{X}$) and divergence-free ($\mathbf{Y}$) vector fields. So this theorem will be our route to the decomposition of $\mathbf{A}$ and $\mathbf{E}$. It guarantees that we can mathematically ``isolate’’ the divergence-free components and curl-free components of $\bf A$ and of $\bf E$. In physics jargon, the divergence-free component is called `transverse’ (also `radiative’), illustrating that it carries two degrees of freedom; and the curl-free component is called `longitudinal’, for one degree of freedom. 

We stress that---quite apart from this paper's aims---although the split of the electric field into radiative and  Coulombic components is mathematically useful,   it is far from being of only mathematical relevance; and from being an arbitrary division of the field's degrees of freedom. The split is physically meaningful. For instance, the question often arises about whether a process `radiates' electromagnetic waves. For example, the famous ``freely-falling electron paradox'' asks this question (see e.g. \cite{Saa_electron} for a pedagogic review). That is because distinguishing the components of the electric field that are wave-like from those which are `Coulombic'  is not straightforward. These are the two characteristics that are expected of any definition of radiation: (i) it is transversal: i.e. it only contains polarizations that are orthogonal to the momenta; (ii) if the charges are confined to a compact region, the Coulombic and radiative parts of the electric field are the coefficients in an inverse radial expansion ($1/R^2$ and $1/R$ terms, respectively).   The radiative $\mathbf{E}^r$ that we will specify below has both these features.

Now, $E_i$ and $A^i$ generate the standard coordinates in phase space. These are coordinates in an infinite-dimensional space, which parametrize six degrees of freedom per spatial point.  But if we want to think of the Gauss law as derivative (as we envisaged in Section \ref{sec:Gauss}), we need to find new coordinates  for phase space (defined in terms of the original coordinates) that instead parametrize the radiative and Coulombic parts of the fields, much like what we did in Section \ref{sec:toy}. The Coulombic coordinate will then be uniquely fixed by the charge density distribution, while the radiative part should parametrize the rest of the field. These coordinates will no longer be local in spacetime, but they are nonetheless very useful, as remarked above. And once we have found these new coordinates, we can use the symplectic form $\Omega$ (cf. \ref{eq:symp_form}) to establish the conjugate decompositions of the gauge potential $\mathbf{A}$.  Indeed, it is easiest to expound this re-definition of coordinates in terms of symplectic geometry, since the symplectic form $\Omega$ is a coordinate-independent object. Moreover, as in Section \ref{sec:toy}, it should be noted that the symplectic form is only non-degenerate in the total phase space. Therefore, to seek a decomposition through symplectic orthogonality, that decomposition should not be restricted to the constraint surface; rather it selects appropriate coordinates in the total phase space, and the constraint surface fixes some of these coordinates. In short: we are seeking adapted coordinates to parametrize the constraint surface. 


Thus, on physical grounds, we would like to decompose the electric field as $E_i = E^r_i + E^c_i$ (with $r$ for `radiative’ i.e. divergence-free, and $c$ for Coulombic, i.e. curl-free), where 
\be\label{eq:E_rad} \pp^i  E^r_i\equiv 0; \qquad\text{and}\qquad \pp^i  E^c_i=\rho.
\ee
Both components are dynamically independent (the evolution of one is independent of the evolution of the other); and $E^c_i$ is to be completely fixed by the distribution of the charges. This means, in vacuum ($\rho=0$), that $ E^c_i=0$ and $E_i=E^r_i$.  In short: ${\bf E}^r$ is the component of the electric field that is not due to the distribution of charges (hence the label `radiative’); while ${\bf E}^c$ is the component of the field due to the simultaneous distribution of charges. What is the form of this component?

The Gauss constraint reduces the degrees of freedom in $\bf E$ from three to two. So when we decompose ${\bf E}$, writing: $E_i = E^r_i + E^c_i$,  with $E^c_i$ to be completely fixed by the distribution of the charges, as discussed above,  we conclude that $E^r_i$ has two degrees of freedom; so that $E^c_i$ has one, i.e. it is `secretly' a scalar. 
 Since one of the terms---$E^r$---is divergence-free, and we want the degrees of freedom of the other---$E^c$---to be exhausted by the divergence, i.e. to be `secretly' a scalar, it is convenient to introduce the Helmholtz decomposition: we take $E^c_i$ to be curl-free, so that it is the gradient of a scalar. 
We  thus \emph{choose} to write this  vector quantity in the customary way, i.e. as  $E^c_i=\pp_i\phi$, for some scalar function $\phi$.\footnote{ By the Poincar\'e lemma, in every simply connected domain, any curl-free vector field---which in the language of differential forms would be written as e.g. $\d \mathbf A=0$---is of a pure gradient form. In the field of tensor algebra, one often refers (confusingly) to a `spin-decomposition'. Thus a 2-tensor $T_{ab}$ may have vector and scalar components, of the form $\nabla_a V_b$ and $\nabla_a\nabla_b\phi$, respectively, and a vector $V_a$ may have a scalar component, that is written as $\nabla_a\phi$. The extraction of these components usually is made through  something like the Helmholtz decomposition theorem.  In fact, the theorem is a special case of the Hodge decomposition theorem for n-forms \cite[Chapter 43]{Morita_book}.} Then, the Gauss constraint $\mathrm{G}=0$ fixes $\phi$  in terms of its simultaneous distribution of charges, via the Poisson equation:
\be\label{eq:Poisson} \nabla^2 \phi_c=\rho. 
\ee

Now, the purely Coulombic terms of the electric field have their phase space coordinate axis  generated by the following vectors on phase space; (recall the discussion around eq. \ref{eq:infv}): 
\be \bb E_c:=\int \d^3 x\,\pp_i\phi \frac{\delta}{\delta E^i(x)} , \quad \text{for each} \quad \phi\in C^\infty(\Sigma).\ee  
This coordinate can then be fixed by \eqref{eq:Poisson}; with $\phi\stackrel{!}{=}\phi_c$ the electrostatic-like potential.\footnote{One might worry here that not all vector fields define a coordinate system: their vector field commutator must vanish, for that to be the case.   For finite dimensions, this is not a worry for $\sum_i a^i \frac{\pp}{\pp x^i}$, since the coefficients are constant, i.e. phase space independent.  Happily, the same is true here: for the coefficients of the new directions do not themselves depend on the coordinates.}

And now we can ask what are the subset of degrees of freedom of $\mathbf{A}$ that are symplectically orthogonal to the Coulombic part of the electric field.  Let us call these components $A^r_i$. That is, we seek those vector fields 
\be\label{eq:Ar}
 \bb A^r=\int \d^3 x A^r_i \frac{\delta}{\delta A_i}\ee 
that are symplectically orthogonal to the Coulombic coordinates of the electric field. That is, the $A^r_i$ are specified by requiring:
\be\label{eq:ortho} 
0\stackrel{!}{=}\Omega( \bb A^r, \bb E_c)=\int \d^3 x\, A^r_i \,\pp^i\phi=-\int \d^3 x\,\phi\,\pp^i A^r_i, \quad \text{for all} \quad \phi\in C^\infty(\Sigma) \, ;
\ee
where we applied integration by parts in the   third  equality and assumed that there is no boundary contribution to the integrals. Since $\phi$ is an entirely arbitrary test (or coordinate) function  we get: 
\be\label{eq:Coulomb_gauge}
\pp^i A^r_i=0. 
\ee
Equation \eqref{eq:Coulomb_gauge} is,   of course,  the \emph{Coulomb gauge} for   (the radiative component of) the gauge potential. It is a complete gauge-fixing, i.e. it leaves no more gauge-freedom in the potential. Moreover,  just like in Section \ref{sec:toy}, where $Q_-$ was written in terms of the original configuration variables, here we could write: 
\be A^r_i(A):=A_i-\pp_i(\nabla^{-2} \pp^j A_j),\ee
which is also called the \emph{radiative projection}. From this equation, it is clear that for any gauge-transformed $A_i^\lambda:=A_i+\pp_i\lambda$, a straightforward computation will show that $A^r_i(A^\lambda)=A^r_i(A)$. That is: $A^r$ is gauge-invariant. 

 Correspondingly, instead of arguing for \eqref{eq:E_rad}  from physical grounds as above, we can see the radiative part of the electric field as being selected   as the component of the electric field that is symplectically orthogonal to the pure gauge part of the gauge potential. Namely, we take the pure gauge part, i.e. the vectors that are along the gauge-orbit, to be given by:
\be\label{eq:Ac} 
\bb A^c:=\int \d^3 x \,\pp_i \lambda \,\frac{\delta}{\delta A_i} , \quad \text{for each} \quad \lambda\in C^\infty(\Sigma) .\ee 
Then, in parallel to the calculation \eqref{eq:ortho}, namely:  
\be 
0\stackrel{!}{=}\Omega( \bb A^c, \bb E_r)=\int \d^3 x\, E^r_i \,\pp^i\lambda=-\int \d^3 x\,\lambda\,\pp^i E^r_i, \quad \text{for all} \quad \lambda\in C^\infty(\Sigma) \, ,
\ee
 we find the defining equation for the fields that are symplectically orthogonal to  the pure-gauge part of $\mathbf{A}$: 
\be\label{eq:Er}
\pp^i E^r_i=0.
\ee

  Thus we have found that if we choose to represent the  Coulombic  degree of freedom by the gradient of a scalar, we obtain the Helmholtz decomposition for the electric field: a unique decomposition in terms of divergence-free and curl-free components. And we obtain a similar decomposition for the gauge potential. That is, we obtain:
\begin{align}
\mathbf{A}=\mathbf{A}^r+\d \lambda\quad\text{and} \quad \mathbf{E}=\mathbf{E}^r+\d \phi.
\end{align}

 To sum up:  we have shown (by just an integration by parts, in each case) that:\\
\indent \indent (1) the curl-free (longitudinal/Coulombic) component of $\bf E$ is symplectically orthogonal to the radiative part of the gauge potential; (cf. equations \eqref{eq:Ar} to \eqref{eq:Coulomb_gauge});\\
while on the other hand:\\
\indent \indent (2) the divergence-free (transverse/radiative) component  of $\bf E$ is symplectically orthogonal to the pure gauge part of gauge field; (cf. equations \eqref{eq:Ac} and \eqref{eq:Er}).

 Finally, we note as a corollary to these results, that we can similarly orthogonally decompose the symplectic form \eqref{eq:symp_form} as:
\be\label{eq:symp_form2}
\Omega  = \int \d^3x  \left(\,\delta E_r^i \curlywedge \delta A^r_i+\delta E_c^i \curlywedge \delta A^c_i\right) \, :
\ee
which guarantees that the respective phase space directions are (symplectically) independent in each summand. \\

\noindent This concludes our main results. We end this Section with three comments:---\\
  \indent \indent (i): In the presence of boundaries,  the determination of $E^r$ does not require the further stipulation of boundary conditions, \emph{if} the gauge-freedom is taken as unconstrained at the boundary (see \cite{Gomes_new, DES_gf, GomesStudies}). Namely, we obtain that, at the boundary $E^r_i n^i=0$, where $\mathbf{n}$ is the vector normal to the boundary.\\
    \indent \indent (ii):   What we have just seen is a general feature of the symplectic geometric treatment. Namely: the tangent space to the constraint surface and the gauge orbits are the symplectic orthogonal complements of each other: cf. Lemma 1.2.2 in \cite{Marsden2007}. Thus  in particular, the radiative part of the electric field is singled out just by the symplectic form and the gauge orbits.\footnote{The general idea, at least in the first-class, or coisotropic, case, is remarkably simple; and thus merits a quick sketch. From \eqref{eq:Ham_Gamma}, vectors $v$ that are tangent to the constraint surface obey $\d\Phi^I(v)=0$ for every $I$. Thus from \eqref{symquant}, we obtain  $\omega(X_I, v)=0$.} \\
  \indent \indent (iii) : We arrived at the Helmholtz decomposition of the gauge potential by considering the symplectic structure and the Gauss constraint. But the Helmholtz decomposition can also be obtained via an inner product structure on the vector bundles in which the fields we are interested in take values. For instance, we split the vector potential into a component along the gauge orbit and one orthogonal to it---its radiative part---according to a  natural metric on the corresponding space of functions. Similarly, according to an inner product on a  different vector bundle on which the gauge group acts, we can ask for such a  generalized Helmholtz decomposition, that splits any given field on a simply connected space into its pure gauge and `radiative', or gauge-orthogonal components. This was studied for general scalar or vector-valued fields in \cite[Section 7]{GomesHopfRiello}, where it was found that the split is only mathematically well-defined where the field is non-vanishing,  in which case it selects unitary gauge.  (For the point-particle analogue of this split along/orthogonal to the gauge orbit, cf. \cite[Section 7]{GomesGryb_KK}.) In those cases, the corresponding fixing of the gauge occurs locally and corresponds to what is labeled in \citep{Francois, Elements_gauge} \emph{an artificial symmetry}: one that can be locally projected out to produce a gauge-invariant ``dressed'' configuration. The dressed state for a nowhere-vanishing Klein-Gordon field corresponds to \cite{Wallace_deflating}'s gauge-invariant, local composite field, with which one can give an entirely local account of the Aharonov-Bohm effect. It is likely, but has not been shown, that similarly to how  in this paper we obtained Coulomb  gauge, we could use the  symplectic structure  to arrive at the unitary  gauge for the Klein-Gordon field.
    

\section{Conclusion and outlook}\label{concl}

Using the symplectic or Hamiltonian formalism, we have shown how decompositions of the electric field correspond, through symplectic orthogonality, to decompositions of the gauge potential (which correspond to choices of gauge), and vice versa. Gauge choices thus have a very natural interpretation in terms of choices of decomposition of the electric field. 

We have argued  that  for various reasons a natural  decomposition of the electric field takes one part, i.e. component,  to be determined by  the instantaneous charge distribution.
We find that  the part of the electric field that remains---i.e. the radiative part, the part that is not involved in the Gauss law and does not ``care about''   the instantaneous charge distribution---is symplectically \textit{orthogonal} to the pure gauge part of the gauge potential. This orthogonality establishes a firm link between gauge symmetry and locality,  as a relation between the Gauss law and the pure gauge part of the potential. But this relation does not, by itself, suffice to select a gauge-fixing.

On the other hand, \emph{if} we define the Coulombic part of the electric field as the gradient of a Coulombic potential, then the gauge \emph{is} fixed:  the  part of the gauge potential that is   symplectically orthogonal to the gradient of the Coulombic potential   \emph{is}  the gauge potential in Coulomb gauge. 

 Thus in summary, again: our main idea is that the electric field has the gauge potential as  its conjugate, and there is a part of the potential that is not at all conjugate to (i.e. is symplectically orthogonal to) the Coulombic part of the electric field, i.e. the part determined by the instantaneous distribution of
charges. {\em That} part of the potential satisfies the Coulomb gauge condition.   And this result is worth expounding, since it shows that a choice of gauge need not be a matter of calculational convenience for some specific problem or class of problems, but can be related  to a physically natural, and general, splitting of the electric field.

 We also saw how these results prompt a comparison with \cite{Maudlin_ontology}. For Maudlin attempted to shift the Gauss law from, in his terms, the nomology to the derivative ontology: namely, by analysing the charges in terms of the electric field. We have argued that this does not work. 
In any case, it does not help with fixing a gauge.  If instead we analyse the Coulombic part of the electric field away, defining it in terms of the charges by \eqref{eq:Poisson}, then  the fundamental ontology  (to use Maudlin's term) can be made to correspond, as we have seen in  \S \ref{sec:radcoul},  to a configuration space parametrised by  the charges and to a choice of the  Coulomb gauge for the gauge potential. 

But we emphasise that  Coulomb gauge is not {\em mandatory}. It only corresponds, in a well-defined sense, to a particular decomposition that singles out the Coulombic part of the electric field. Nonetheless, in whichever sense that choice of decomposition of the electric field is natural, Coulomb gauge is also natural. 

\medskip

Finally, we offer an {\em outlook}: we stress that the lessons of this paper go through, with minor modifications, to the non-Abelian domain,  and also apply in the presence of boundaries; (see e.g. \cite{GomesStudies, GomesRiello_new}). In brief, each extension requires one important modification. 

First, in the non-Abelian case, the Coulombic split occurs only at the level of \emph{perturbations}. That is, although we can split vectors on phase space, $\bb X$,  the non-Abelian nature of the theory implies that the split is not integrable (see \cite[Section 5]{GomesRiello_new} and \cite[Section 9]{GomesHopfRiello}). Thus the Coulombic split of a state is always ambiguous.\footnote{In the Abelian case, one can integrate the perturbative split along paths in phase space so as to define a split of any final state; integrability guarantees that the end result is path-independent.}  

In the presence of boundaries, in the Abelian case, the radiative and Coulombic components of the electric field are again defined by symplectic orthogonality with the pure gauge part of the potential, which we deem unconstrained at the boundary (cf. \cite[Prop. 3.3]{GomesRiello_new}). Thus we obtain the following modifications, for a bounded region $R$ bounded by $\pp R$, whose normal is $\mathbf{n}$:\\
Instead of \eqref{eq:Coulomb_gauge}, 
\be
\begin{dcases}
\pp^i A^r_i=0  & \text{in }R,\\
 n^iA^r_i=0  & \text{at }\pp R.
\end{dcases};
\ee  
instead of \eqref{eq:Er},
\be
\begin{dcases}
\pp^i E^r_i=0 & \text{in }R,\\
 n^iE^r_i=0  & \text{at }\pp R.
\end{dcases};
\ee 
and instead of \eqref{eq:Poisson},
\be
\begin{dcases}
\nabla^2\phi_c=\rho  & \text{in }R,\\
 n^i\pp_i\phi=f & \text{at }\pp R.
\end{dcases},
\ee
where $f$ is the electric flux through the boundary, $f:=n^i E_i$, and here can be seen as an independent variable. Note that: (i) the pure gauge part of the potential is unmodified (since we do not truncate gauge transformations at the boundary $\pp R$), and (ii) the normal to the electric field at the boundary belongs to the Coulombic part. The radiative part gets no extra degree of freedom at the boundary.\footnote{Thus, the radiative part is local, according to straightforward definitions of locality (cf.  \cite[\S 2, p.5]{Wallace2019b}).  The Coulombic field depends on the distribution of $\rho$ in the region \emph{and} on the boundary flux of the electric field. Thus, even if the charge distribution for two worlds matches inside $R$, the Coulombic field therein may differ, since in each world it will depend on the independent variable that is the electric boundary flux.  }

\subsection*{Acknowledgements}

Many thanks to Caspar Jacobs for helpful comments on an earlier version of this paper.

\begin{appendix}

\section*{Appendix}

\section{Holonomies and non-separability}\label{app:hol}

The holonomy interpretation of electromagnetism takes as its basic elements assignments of unit complex numbers to loops in spacetime. A loop is the image of a smooth  embedding of the oriented circle,  $\gamma:S^1\rightarrow \Sigma$; the image is therefore a closed, oriented, non-intersecting curve. One can form a basis of gauge-invariant quantities for the holonomies (cf. \cite{Barrett_hol} and \cite[Chapter 4.4]{Healey_book} and references therein):\footnote{Of course, any discussion of matter charges and normalization of action functionals would require $e$ and $\hbar$ to appear. However, we are not treating matter, so these questions of  choice of unit do not become paramount. As before, if needed, we set our units  to $e=\hbar=1$; as is the standard choice in quantum chromodynamics (or as in the so-called Hartree convention for atomic units).} 
\be\label{eq:hol} 
hol(\gamma):=\exp{(i\int_\gamma A)}.
\ee
In more detail: by exponentiation (path-ordered in the non-Abelian case), we can assign a complex number  (matrix element in the non-Abelian case) $hol(C)$ to the oriented embedding of the unit interval: $C:[0,1]\mapsto M$. This makes it easier to see how composition works: if the endpoint of $C_1$ coincides with the starting point of $C_2$, we define the composition $C_1\circ C_2$ as, again, a map from $[0,1]$ into $M$, which takes $[0,1/2]$ to traverse $C_1$ and $[1/2, 1]$ to traverse $C_2$.  The inverse $C^{-1}$ traces out the same curve with the opposite orientation, and therefore $C\circ C^{-1}=C(0)$.\footnote{Intuitively, we do not want to consider curves that trace the same path back and forth, i.e.  \textit{thin} curves. Thus we  define a closed curve as \textit{thin} if it is possible to shrink it down to a point while remaining within its image. Quotienting the space of curves by those that are thin, we obtain the space of \textit{hoops}, and this is the  space actually considered in the treatment of holonomies.  We will not emphasise this finer point, since it follows from an intuitive understanding of the composition of curves.} 
Following this composition law, it is easy to see from \eqref{eq:hol} that 
\be\label{eq:loop_com} hol(C_1\circ C_2)=hol(C_1)hol(C_2),\ee with the right-hand-side understood as complex multiplication in the Abelian case, and as composition of linear transformations, or  multiplication of matrices, in the non-Abelian case.

As both Healey \cite[Chapter 4.4]{Healey_book} and Belot (\cite[Section 12]{Belot2003} and \cite[Section 3]{Belot1998}) have pointed out: even classical electromagnetism, in the holonomy interpretation, exhibits a form of non-locality, which   at first sight, one might  have thought was a hallmark of non-classical physics.


But one naturally asks: does the state of a region nevertheless supervene on assignments of intrinsic properties to the sub-regions of the region (where the sub-regions  may be
taken to be arbitrarily small)? This is essentially the question whether the theory is \textit{separable}: (see \cite[Chapter 2.4]{Healey_book}, \cite[Section 3]{Belot1998}, \cite[Section 12]{Belot2003},  \cite{sep-physics-holism, Myrvold2010}    and  \cite[Section 2, p.5]{Wallace2019b}).  For this topic, we can focus on Myrvold's definition \cite[p.427]{Myrvold2010}. (It builds on Healey's notion of Weak Separability \cite[p. 46]{Healey_book} and on Belot's notion of Synchronic
Locality \cite[p. 540]{Belot1998}.)\\

\begin{figure}[t]
		\begin{center}
			\includegraphics[width=8cm]{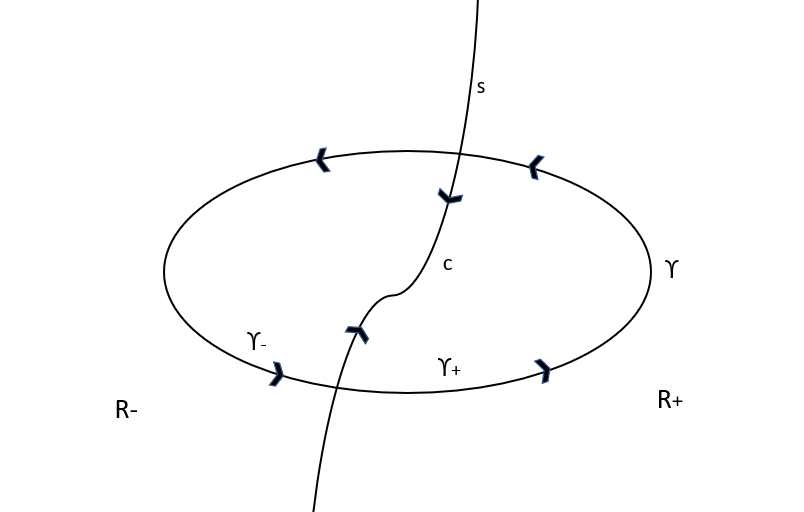}	
			\caption{ Two subregions, i.e. $R_\pm$,  with the separating surface $S$. A larger loop $\gamma$ crosses both regions. But, since $\gamma_1$ and $\gamma_2$ traverse $S$ along $C$ in opposite directions, $\gamma=\gamma_1\circ\gamma_2$.
			}
			\label{fig4}
		\end{center}
\end{figure}

\noindent$\bullet$ \textit{Patchy Separability for Simply Connected Regions}. For any simply connected spacetime region $R$, there are arbitrarily fine open coverings $\mathcal{N}=\{R_i\}$ of $R$ such that the state of $R$ supervenes on an assignment of qualitative intrinsic physical properties to elements of $\mathcal{N}$.\\

For electromagnetism in vacuum,   it is easy to show that \textit{Patchy Separability for Simply Connected Regions}  \textit{does} hold. In Figure 2, we see a loop $\gamma$ not contained in either $R_+$ or $R_-$. However, we can decompose it as $\gamma=\gamma_+\circ\gamma_-$, where each regional loop $\gamma_\pm$ does not enter the complementary region ($R_\mp$, respectively), but each traverses the  curve $c$ lying within the boundary $S$ each in the opposite direction to the other. Using \eqref{eq:loop_com},  since the contributions to the holonomies of the two paths along $c$ cancel out, the fact that holonomies form a basis of gauge-invariant quantities,  then implies  that  the gauge-invariant state of the whole  region supervenes on  the gauge-invariant states of its sub-regions. As one might put it: the gauge-invariant state of the universe, according to the theory, supervenes on the  gauge-invariant states of its regions. This  corresponds to the fact that in the absence of charges, and for a simply connected manifold, we can solve for the Coulombic part of the field simply by setting $E_c=0$.

 But Patchy Separability fails for non-simply connected regions. This is easy to see from Figure 2: if one introduces a hole in the middle of the curve $c$, say composed of two semi-circles $c_1$ and $c_2$ that meet along $c$,  we can no longer write $\gamma=\gamma_+\circ\gamma_-$, since curves in each region, traversing either $c_1$ or $c_2$,  are not aligned and thus their contribution does not cancel out. (Compare  \cite[Section 6.8.2]{GomesRiello_new} for how this relates to Coulombic vs. radiative modes, and to the topological non-locality of the Aharonov-Bohm effect.).

It is also easy to see how \textit{Patchy Separability for Simply Connected Regions} {\em fails} when charges are present within the regions but absent from the boundary $S$;  (see in particular  \cite[Section 4.3.2]{GomesRiello_new}, and footnote 70 in \cite{Gomes_new}). For, in the presence of charges, we can form \textit{gauge-invariant} functions from a non-closed curve $C'$  that crosses $S$ and has  e.g. one positive and one negative charge, $\psi_\pm(x_\pm)$, capping off $C'$ at $x_\pm\in R_\pm$. That is, the following quantity is a gauge-invariant function:
$$Q(C', \psi_\pm)= \psi_-(x_-)hol(C')\psi_+(x_+)
$$
for $C'(0)=x_-, C'(1)=x_+$. It is easy to check from the transformation  properties, $A_i\mapsto A_i+\pp_i\lambda$ and    $\psi\mapsto e^{i\lambda}\psi$, that $Q$ is gauge-invariant. Moreover, we cannot break this invariant up into   gauge-invariant contributions from the two regions, since we have assumed no charges lie at the boundary. 

\end{appendix}

\end{document}